\documentclass[12pt]{article}
\usepackage[usenames]{color}
\usepackage{framed,amsmath,amssymb,euscript,cite,amsfonts,graphicx}
\usepackage{pstricks,fancybox,hyperref}
\usepackage[all,knot,poly]{xy}
\usepackage{wrapfig}

\fontfamily{pnc}\selectfont

\definecolor{shadecolor}{rgb}{0.7,.5,.3}

\begin{document}
\setlength{\parskip}{2ex} \setlength{\parindent}{0em}
\setlength{\baselineskip}{3.6ex}
\newcommand{\onefigure}[2]{\begin{figure}[htbp]
         \caption{\small #2\label{#1}(#1)}
         \end{figure}}
\newcommand{\onefigurenocap}[1]{\begin{figure}[h]
         \begin{center}\leavevmode\epsfbox{#1.eps}\end{center}
         \end{figure}}
\renewcommand{\onefigure}[2]{\begin{figure}[htbp]
         \begin{center}\leavevmode\epsfbox{#1.eps}\end{center}
         \caption{\small #2\label{#1}}
         \end{figure}}
\newcommand{\comment}[1]{}
\newcommand{\myref}[1]{(\ref{#1})}
\newcommand{\secref}[1]{sec.~\protect\ref{#1}}
\newcommand{\figref}[1]{Fig.~\protect\ref{#1}}
\def\sl2z{SL(2,\Z)}
\newcommand{\be}{\begin{equation}}
\def\bp{\,{\hbox{P}\!\!\!\!\!\hbox{I}\,\,\,}}
\def\br{\,{\hbox{R}\!\!\!\!\!\hbox{I}\,\,\,}}
\def\bc{\,{\hbox{C}\!\!\!\hbox{I}\,\,\,}}
\def\bz{\,{\hbox{Z}\!\!\hbox{Z}\,\,\,}}
\def\lk{{\rm lk}}
\def\a{\alpha}
\newcommand{\ee}{\end{equation}}
\newcommand{\bea}{\begin{eqnarray}}
\newcommand{\eea}{\end{eqnarray}}
\newcommand{\beqa}{\begin{Beqnarray*}}
\newcommand{\eeqa}
{\end{Beqnarray*}}
\newcommand{\nn}{\nonumber}
\newcommand{\unit}{1\!\!1}
\newcommand{\q}{q_{1}}
\newcommand{\qq}{q_{1}^{2}}
\newcommand{\qqq}{q_{1}^{3}}
\newcommand{\qqqq}{q_{1}^{4}}
\newcommand{\p}{q_{2}}
\newcommand{\pp}{q_{2}^{2}}
\newcommand{\ppp}{q_{2}^{3}}
\newcommand{\pppp}{q_{2}^{4}}
\newcommand{\R}{\bf R}
\newcommand{\X}{{\bf X}}
\newcommand{\T}{{\bf T}}
\newcommand{\PP}{\bf P}
\newcommand{\CC}{\bf C}
\newcommand{\CJ}{\mathcal{J}}
\def\Z{{\bf Z}}
\def\aa{{\bf a}}
\newdimen\tableauside\tableauside=1.0ex
\newdimen\tableaurule\tableaurule=0.4pt
\newdimen\tableaustep
\def\phantomhrule#1{\hbox{\vbox to0pt{\hrule height\tableaurule width#1\vss}}}
\def\phantomvrule#1{\vbox{\hbox to0pt{\vrule width\tableaurule height#1\hss}}}
\def\sqr{\vbox{%
\phantomhrule\tableaustep
\hbox{\phantomvrule\tableaustep\kern\tableaustep\phantomvrule\tableaustep}%
\hbox{\vbox{\phantomhrule\tableauside}\kern-\tableaurule}}}
\def\squares#1{\hbox{\count0=#1\noindent\loop\sqr
\advance\count0 by-1 \ifnum\count0>0\repeat}}
\def\tableau#1{\vcenter{\offinterlineskip
\tableaustep=\tableauside\advance\tableaustep by-\tableaurule
\kern\normallineskip\hbox
    {\kern\normallineskip\vbox
      {\gettableau#1 0 }%
     \kern\normallineskip\kern\tableaurule}%
  \kern\normallineskip\kern\tableaurule}}
\def\gettableau#1 {\ifnum#1=0\let\next=\null\else
  \squares{#1}\let\next=\gettableau\fi\next}

\tableauside=1.0ex \tableaurule=0.4pt

\setcounter{page}{1} \pagestyle{plain}

\definecolor{myorange}{rgb}{1,0.5,0}
\definecolor{darkslategray}{rgb}{0.18431,0.3098,0.3098}
\definecolor{darkslategrey}{rgb}{0.18431,0.3098,0.3098}
\definecolor{dimgray}{rgb}{0.41176,0.41176,0.41176}
\definecolor{dimgrey}{rgb}{0.41176,0.41176,0.41176}
\definecolor{slategray}{rgb}{0.43921,0.50195,0.5647}
\definecolor{slategrey}{rgb}{0.43921,0.50195,0.5647}
\definecolor{lightslategray}{rgb}{0.46666,0.53333,0.59999}
\definecolor{lightslategrey}{rgb}{0.46666,0.53333,0.59999}
\definecolor{gray}{rgb}{0.74509,0.74509,0.74509}
\definecolor{grey}{rgb}{0.74509,0.74509,0.74509}
\definecolor{lightgrey}{rgb}{0.82744,0.82744,0.82744}
\definecolor{lightgray}{rgb}{0.82744,0.82744,0.82744}

\begin{titlepage}
\begin{center}

\hfill hep-th/yymmnnn\\ \vskip 1cm

{\Large{\bf { Link Homologies and the Refined Topological Vertex}}} \vskip
0.5cm
{\large {\sc Sergei Gukov$^{1}$,\,\,Amer Iqbal$^{2}$\,,\,\,Can Koz\c{c}az$^{3}$\,,\,\,Cumrun Vafa$^{4,5}$}\\
\vskip 0.5cm
{{$^{1}$Department of Physics,
University of California,\\
Santa Barbara, CA, 93106\\} \vskip 0.2cm {$^{2}$ Department of Mathematics, University of Washington,\\
Seattle, WA, 98195\\}\vskip 0.2cm
{$^{3}$Department of Physics,
University of Washington,\\
Seattle, WA, 98195\\}\vskip 0.2cm
{$^{4}$Center for Theoretical Physics,
Massachusetts Institute of Technology,\\
Cambridge,  MA, 02139\\} \vskip 0.2cm {$^{5}$Jefferson Physical Laboratory, Harvard University,\\
Cambridge, MA, 02138}}}
\end{center}
\vskip 0.7 cm
\begin{abstract}\baselineskip 0.63cm
{\large We establish a direct map between refined topological vertex
and $sl(N)$ homological invariants of the of Hopf link, which include
Khovanov-Rozansky homology as a special case.
This relation provides an exact answer for homological invariants
of the of Hopf link, whose components are colored by arbitrary
representations of $sl(N)$. At present, the mathematical formulation
of such homological invariants is available only for the fundamental
representation (the Khovanov-Rozansky theory) and the relation with
the refined topological vertex should be useful for categorifying
quantum group invariants associated with other representations $(R_1, R_2)$.
Our result is a first direct verification of a series of conjectures
which identifies link homologies with the Hilbert space of BPS
states in the presence of branes, where the physical interpretation
of gradings is in terms of charges of the branes ending on Lagrangian branes.}

\end{abstract}

\end{titlepage}

{\small \tableofcontents}
\baselineskip 0.63cm

\section{\sc Introduction}\label{introduction}
One of the most promising recent developments in a deeper
understanding of link invariants involves the study of homological
invariants. First, these invariants provide a refinement of the
familiar polynomial invariants. Secondly, and more importantly, they
often lift to functors. However, constructing such homological
invariants for arbitrary groups and representations has been a
challenging problem, and at present only a handful of link
homologies is known. Most of the existing examples are related to
the fundamental representation of classical groups of type $A$ and
include the Khovanov homology \cite{Khovanov}, the link Floer
homology \cite{OShfk,OShfl,Rasmussen}, and the $sl(N)$ knot homology
\cite{Khovanoviii,RKhovanov}.

On the physics side, polynomial invariants of knots and links
can be realized in the Chern-Simons gauge theory \cite{WittenJones}.
On the other hand, a physical interpretation of link homologies was
first proposed in \cite{GSV} and further developed in \cite{DGR,GWalcher}.
The interpretation involves BPS states in the context
of physical interpretation of open topological string amplitudes \cite{OV}.
In order to explain the realization in topological string
theory one first needs to consider embedding the Chern-Simons gauge theory
in string theory \cite{Wittencsstring} and the large $N$ dual
description in terms of topological strings \cite{GopakumarV}.
As was shown in \cite{OV} and will be reviewed in the next section,
in this dual description polynomial invariants of knots
and links are mapped to open topological string
amplitudes which, in turn, can be reformulated in terms of
integer enumerative invariants counting degeneracy of states in Hilbert spaces,
roughly the number of holomorphic branes ending on Lagrangian branes.
This leads to a physical reformulation of polynomial link invariants
in terms of the so-called Ooguri-Vafa invariants which, roughly
speaking, compute the Euler characteristic of the $Q$-cohomology,
that is cohomology with respect to the nilpotent components of
the supercharge\footnote{Elements of this cohomology can be viewed
as the ground states of the supersymmetric theory of M2 branes ending
on M5 branes in a particular geometry \cite{OV}, as we review below.}.

\begin{table}\begin{center}
\begin{tabular}{|c||c|c|c|}
\hline
\rule{0pt}{5mm}
& Rational & Integer & Refinement \\[3pt]
\hline
\hline
\rule{0pt}{5mm}
Closed & Gromov-Witten & Gopakumar-Vafa/ & Refined BPS invariants \\[3pt]
&  & Donaldson-Thomas &  \\[3pt]
\cline{1-4}
\rule{0pt}{5mm}
Open & open & Ooguri-Vafa & triply graded invariants \\[3pt]
 & Gromov-Witten & invariants & $D_{J,s,r}$ and $N_{J,s,r}$ \\[3pt]
\hline
\end{tabular}\end{center}
\caption{Enumerative invariants of Calabi-Yau three-folds.}
\end{table}

This, however, is not the full answer to homological link invariants
which require the understanding of an extra grading.
In other words, there is an extra physical charge needed
to characterize these invariants.
In closed string theory, an extension of topological
string was constructed for certain non-compact Calabi-Yau
geometries \cite{Nekrasov}. It involves an extra parameter
which has the interpretation of an extra rotation in
the four-dimensional space.  It was shown in \cite{HIV}
that this extra charge indeed accounts for the charges
of the M2 branes on holomorphic curves inside a Calabi-Yau three-fold.

It was proposed in \cite{GSV} that the homological grading
of link homologies is related to the extra charge in
the extension of topological string proposed in \cite{Nekrasov}.
In particular, supersymmetric states of holomorphic branes
ending on Lagrangian branes, labeled by all physical charges,
should reproduce homological invariants of knots and links,
\bea
{\cal H} (L) = {\cal H}_{BPS}
\label{hhbps}
\eea
This conjecture led to a number of predictions regarding
the structure of $sl(N)$ knot homologies, in particular
to the triply-graded knot homology categorifying
the HOMFLY polynomial \cite{DGR, RKhovanovii}, see also \cite{GWalcher}.
However, 
a direct test of this conjecture and computation of homological link
invariants from string theory was difficult due to lack of
techniques suitable for calculating degeneracies of BPS states
in the physical setup.

Thus, even for the unknot, the only case where one can compute both
sides of \eqref{hhbps} independently is the case of the fundamental
representation. For other representations, a mathematical
formulation of homological knot invariants is not available at
present, while on the string theory side the direct analysis of
${\cal H}_{BPS}$ becomes more difficult. For a certain class of
representations --- which, for example, include totally symmetric
and totally anti-symmetric representations of $sl(N)$ --- it was
argued in \cite{GWalcher} that the corresponding cohomology ring of
the unknot, ${\cal H}^{{\bf g}, R}$, is related to the Jacobi ring
of a potential $W_{{\bf g}, R} (x_i)$,
\bea
{\cal H}^{{\bf g}, R} ({\rm unknot}) \cong \CJ ( W_{{\bf g}, R} (x_i) )
\label{hwunknot}
\eea
It is expected that for this class of representations the
corresponding link homologies can be defined using matrix
factorizations of the potential $W_{{\bf g}, R} (x_i)$, as in the
original construction of the Khovanov and Rozansky \cite{RKhovanov}.
The simplest set of examples of such representations involves
totally anti-symmetric representations of $sl(N)$. For the the
$k$-th antisymmetric representation of $sl(N)$, the potential is the
Landau-Ginzburg potential of $A_N^{\otimes k}$ minimal model, and
the corresponding homology ring of the unknot \eqref{hwunknot} is
the cohomology ring of the Grassmannian of $k$-planes in
$\mathbb{C}^N$ \cite{RKhovanov,GWalcher},
\bea
{\cal H}^{ sl(N) , \Lambda^k} ({\rm unknot}) \cong H^* (Gr (k,N))
\label{grassmh}
\eea
where all cohomology groups are localized in the single homological
grading. This will be one of our examples below.

 We will be able to compute the homology groups ${\cal H}^{{\bf g}, R}$
directly from string theory using the recent work \cite{IKV}, where
it was shown how the topological vertex \cite{AKMV} (which computes
topological string amplitudes in toric geometries) can be refined to
compute Refined BPS invariants \cite{HIV}. Since the topological
vertex formalism is composed of open string amplitudes, this
refinement together with the conjecture of \cite{GSV} implies that
the refined topological vertex should be computing homological link
invariants, at least for the class of links which can be formulated
in terms of local toric geometries. The basic example of such link
is the Hopf link. This is one of the few examples where we can
directly verify our conjectures, at least in the case of the
fundamental representation, where Khovanov-Rozansky homology of the
Hopf link can be computed. We find in this paper that these highly
non-trivial computations agree with each other exactly!

This provides a strong check of the various conjectures leading to
this statement. Moreover, since the refined topological vertex is
easily computable for arbitrary representations, this leads to a
prediction of {\it all} homological invariants of a large class of
links (of which the Hopf link is the simplest example) colored by
arbitrary representations $(R_1, \ldots, R_{\ell})$,
\bea
{\mathcal H}^{sl(N); R_1, \ldots, R_{\ell}} (L)
\label{hrrrrr}
\eea
This is a highly non-trivial new prediction which we are currently studying,
and it would be very interesting to compare it with the mathematical
formulation of link homologies, once those are developed.
It is likely that these predictions lead to a deeper
mathematical understanding of homological link invariants.
In particular, we hope that the combinatorial interpretation
of the refined vertex in terms of 3D partitions will be useful
for finding combinatorial definition of link homologies \eqref{hrrrrr}.

The organization of this paper is as follows: In section \ref{sec2}
we review the relation between the BPS state counting, link
invariants, and open topological strings, including the large $N$
description of the Chern-Simons theory. In section \ref{sec3} we
review aspects of homological link invariants and their
interpretation as Hilbert spaces of BPS states. In particular, we
use this interpretation to compute the Khovanov-Rozansky homology of
the Hopf link. In section \ref{sec4} we review the refined
topological vertex, which is used in section \ref{sec5} --- together
with some facts from section \ref{sec2} --- to compute the
homological invariants for the Hopf link colored by
arbitrary representations $(R_1,R_2)$, see Eq.\eqref{hopflm} below.
In particular, in the case of the fundamental representation we reproduce the
Khovanov-Rozhansky homology derived in section \ref{sec3}, and make
new predictions.

{\bf Conventions:} The triply-graded invariants discussed in this
paper are naturally organized into generating functions, which are
polynomials in three variables. Unfortunately, the conventions
between the physics literature and the knot theory literature are
slightly different. In order to be careful about such differences
and to agree with the standard notations, we use the variables
$(Q, q_1, q_2)$ when we talk about topological string amplitudes
computed by the topological vertex, {\it cf.} \cite{IKV}, and we use
the variables $({\bf a}, q, t)$ when we discuss link homologies,
{\it cf.} \cite{DGR}. The two sets of variables are related as follows:
\bea
\label{resultt}
\sqrt{q_{2}}&=&\,q\\\nn
\sqrt{q_{1}}&=&\,-t\,q\\\nn
Q&=&\,-t\,{\bf a}^{-2}\,.
\eea
In particular, expressions written in terms of $({\bf a}, q, t)$
involve integer powers of $q$ and $t$, while expressions written
in terms of $(Q, q_1, q_2)$ involve half-integer powers of $q_1$ and $q_2$.
Specialization to the Ooguri-Vafa invariants and to knot polynomials
is achieved, respectively, by setting $q_1 = q_2$ and $t=-1$.

\section{\sc BPS States, Link Invariants, and Open Topological Strings}
\label{sec2}

For the benefit of the reader not very familiar with the description
of D-branes in toric varieties, following \cite{HV,AKV}, let us briefly
review the basics of this description necessary for understanding
the topological string interpretation of link homologies.
Consider a toric variety,
\bea
X = {\bf C}^{k+3}/U(1)^k
\label{xquot}
\eea
where ${\bf C}^{k+3}$ is parametrized by
coordinates $X^i$, $i=1, \ldots, k+3$,
and the symplectic quotient is obtained by imposing
\begin{eqnarray}
&& D^a
= Q_1^a |X^1|^2 + Q_2^a |X^2|^2 + \ldots + Q_{k+3}^a |X^{k+3}|^2 - r^a = 0 \cr
&& U(1)_a ~:~~~ X^i \to e^{i Q_i^a \epsilon_a} X^i
\end{eqnarray}
for every $a=1, \ldots, k$.
We can think of (\ref{xquot}) as a gauged linear sigma model
with gauge group $U(1)^k$ and chiral fields $X^i$ of charges $Q_i^a$.
The charges $Q_i^a$ should obey
$$
\sum_i Q_i^a = 0
$$

Using toric geometry, we can also describe Lagrangian D-branes
invariant under the torus action. There are two interesting types
of Lagrangian D-branes:

\begin{enumerate}

\item
Lagrangians, which project to a 1-dimensional subspace in
the base of the toric variety $X$. These can be described
by three equations of the form
\begin{eqnarray}
&& \sum_i q_i^{\alpha} |X_i|^2 = c^{\alpha}, \quad\quad \alpha=1,2 \cr
&& \sum_i \arg X_i = 0
\end{eqnarray}
where $q_i^{\alpha}$ is a set of charges such that $\sum_i q_i^{\alpha}=0$.

\item
Lagrangians, which project to a 2-dimensional subspace in
the base of the toric variety $X$.
These can be defined by the following equations
\begin{eqnarray}
&& \sum_i q_i^1 |X_i|^2 = c, \cr
&& \sum_i q_i^{\alpha} \arg X_i = 0 \quad\quad \alpha=2,3
\end{eqnarray}
where the charges
should satisfy $\sum_i q_i^1 q_i^{\alpha}=0$, $\alpha=2,3$

\end{enumerate}

\begin{figure}
\begin{center}
\includegraphics[width=2in]{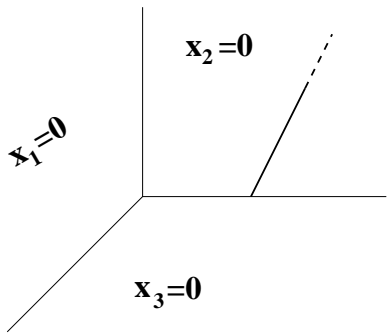}\\
\end{center}
\caption{A Lagrangian D-brane in ${\bf C}^3$, projected to the base
of the toric fibration.} \label{slagfig}
\end{figure}

Let us consider $X= {\bf C}^3$ with a Lagrangian
D-brane on $L$, where $L$ is defined by
\begin{eqnarray}
&& |X_1|^2 - |X_3|^2 = c >0 \cr
&& |X_2|^2 - |X_3|^2 = 0 \cr
&& \sum_i \arg X_i = 0
\label{lone}
\end{eqnarray}
The projection of this Lagrangian D-brane to the base of toric
fibration is shown on Figure \ref{slagfig}.

\subsection{\sc Geometric Transition and the Hopf link}

The conjecture on the geometric transition \cite{GopakumarV} was
originally checked at
the level of free energies and later at the
level of observables of the theory in more detail in \cite{OV}.
A worldsheet explanation of this duality was discovered in \cite{Ooguri:2002gx}. See \cite{marino} for a detailed review of this duality and its consequences for link invariants.

 Let
us briefly review the conjectured equivalence between the
Chern-Simons theory in ${\mathbf S}^{3}$ with the closed topological
string theory on the resolved conifold, or in other words, with the
open topological string theory on $T^{*}{\mathbf S}^{3}$.

In his work, 't Hooft noted that $U(N)$ or $SU(N)$ gauge theories
should have a string theory description. If we consider the
perturbative Feynman diagram expansion in the 't Hooft coupling
$\lambda=N g$ using the double line notation these diagrams
can be regarded as a triangulation of a Riemann surface. The
contributions to the free energy coming from these diagrams can be
arranged in a way that looks like open string expansion on
worldsheet with genus $g$ and $h$ boundaries:
\begin{equation}\label{hooft}
F=\sum_{g=0,h=1}C_{g,h}N^{2-2g}\lambda^{2g-2+h}
\end{equation}
It was shown by Witten for the $SU(N)$ Chern-Simons theory on a
three dimensional manifold ${\mathbf S}^{3}$ that the coefficients
$C_{g,h}$ are equal to the A-model topological open string theory on
a worldsheet with genus $g$ and $h$ boundaries \cite{Wittencsstring}
with the target space $T^{*}{\mathbf S}^{3}$. The $N$ D-branes are
wrapped on the base ${\mathbf S}^{3}$ in this six dimensional
cotangent bundle. The summation over the number of holes in the
Eq.\ref{hooft} can be carried out first. The free energy takes the
following form which looks like the closed string expansion.
\begin{equation}
F=\sum_{g=0}N^{2-2g}F_{g}(\lambda)
\end{equation}
where $\lambda$ acts like some modulus of the theory. The natural
question that arises is ``what is the closed string theory for the
Chern-Simons theory on ${\mathbf S}^{3}$ ?'' In \cite{GopakumarV} it
was conjectured that if we start with the open topological string theory
on $T^{*}{\mathbf S}^{3}$ which can be regarded as the deformed
conifold and wrap $N$ D-branes on the base and take the large $N$
limit, the geometry of the target space undergoes the conifold
transition: the base ${\mathbf S}^{3}$ shrinks and then is blown up
to ${\mathbf S}^{2}$, where the D-branes disappear.  Instead, The
K\"{a}hler moduli of the blown up
${\mathbf S}^{2}$ is proportional to the 't Hooft coupling. The
equivalence was checked for all values of the 't Hooft coupling and
for all genera of the free energy of the Chern-Simons theory and the
closed topological strings on the resolved conifold.

It is worth mentioning that the resolution of the geometry, however,
is not unique:  two different ways of resolving the singularity give rise
to topologically distinct spaces which are birationally equivalent.
In \figref{flop}, two different resolutions of the conifold
singularity are shown which are related by flop. If we insert probe
branes in the target geometry and compute the open string partition
function using the ``usual'' topological vertex the partition
function is invariant under flop. However, for the ``refined''
topological vertex this invariance does not hold, and it will be
crucial in our discussion to choose the `correct' blowup.

\subsection{\sc Knots, links and open topological string amplitudes}
The equivalence  between the open topological string on the
deformed conifold and the closed string on the resolved conifold
was also checked in terms of the observables \cite{OV}. The basic
observables
in the Chern-Simons theory are the Wilson loops. As mentioned
before, there are $N$ D-branes wrapped on the base, and to study
their dynamics another set of D-branes can be introduced, say $M$ of
them. This new set of D-branes will be wrapped on a Lagrangian
3-cycle which is associated with a knot. A closed loop $q(s)$,
$(0\leq s<2\pi)$, is used to parametrize a knot in ${\mathbf
S}^{3}$. Then the conormal bundle associated with the knot defined
as
\begin{equation}
\mathcal{C}=\left\{(q(s),p)\,|\,\,
p^{i}\frac{dq_{i}}{ds}=0,\,\,0\leq s<2\pi \right\}
\end{equation}
is Lagrangian. The $M$ D-branes wrapped on the Lagrangian cycle
$\mathcal{C}$ gives rise to $SU(M)$ Chern-Simons theory. However, in
addition to the Chern-Simons theory on $\mathcal{C}$ there is
another topological open string sector coming from strings stretching
between the $M$ D-branes around $\mathcal{C}$ and the $N$ D-branes
around the base ${\mathbf S}^{3}$. We obtain a complex scalar which
transforms as bi-fundamental of $SU(N)\otimes SU(M)$
and lives in the intersection of the
D-branes, i.e. on the knot. This complex field can be integrated out
and we obtain an effective action for the $U(N)$ gauge connection
$A$ on ${\mathbf S}^{3}$
\begin{equation}
S_{CS}(A)+\sum_{n=1}^{\infty}\frac{1}{n}Tr U^{n} Tr V^{-n}.
\end{equation}
which can be rephrased as correlations of \cite{LMtorus}
\begin{equation}
\langle \sum_R Tr_R U Tr_R V^{-1} \rangle.
\end{equation}
In the previous section we metioned that the geometry changes from
deformed conifold with branes to the resolved conifold without
branes if we take the large $N$ limit. We can take the same limit in
this brane system while keeping the number of non-compact probe
branes, $M$, fixed and trace what happens to the probe branes during
this transition. According to \cite{OV}, the non-compact Lagrangian
cycle $\mathcal{C}$ will be mapped to new Lagrangian cycle
$\mathcal{C'}$ in the resolved conifold, with $M$ D-branes wrapping
it. This will provide boundary conditions for the open strings to end on
in the resolved geometry.  Aspects of this transition including how
one can find the Lagrangian brane for certain knots and links
(including the Hopf link) have been discussed in
\cite{Marino:2001re}.   Precise mathematical description of the
Lagrangian D-brane $\mathcal{C'}$ after transition has been offered
\cite{Taubes}.

For the case of the unknot, discussed in detail in \cite{OV}, the
normalized CS expectation is given by \bea W_{\lambda(R)}=\langle
\mbox{Tr}_{R}U\,\rangle\,,\,\,\,\,\,U=Pe^{\oint A} \eea Where
$\lambda(R)$ is the highest weight of the irreducible representation
$R$ i.e., it is a 2D partition. The above expectation value can be
calculated exactly and is given by
 \bea W_{\lambda}=\mbox{Quantum
dimension of $\lambda$}&=&\prod_{(i,j)\in
\lambda}\frac{q_{1}^{\frac{N+c(i,j)}{2}}-q_{1}^{-\frac{N+c(i,j)}{2}}}{q_{1}^{\frac{h(i,j)}{2}}-q_{1}^{-\frac{h(i,j)}{2}}}\\\nn
&=&q_{1}^{-N\frac{|\lambda|}{2}}\,s_{\lambda}(1,q_{1},q_{1}^2,\cdots,q_{1}^{N-1})
\eea Where $s_{\lambda}({\bf x})$ is the Schur function labelled by
the partition $\lambda$ and $c(i,j)=j-i$,
$h(i,j)=\nu_{i}-j+\nu^{t}_{j}-i+1$ are the content and the hook
length of a box in the Young diagram of $\lambda$.

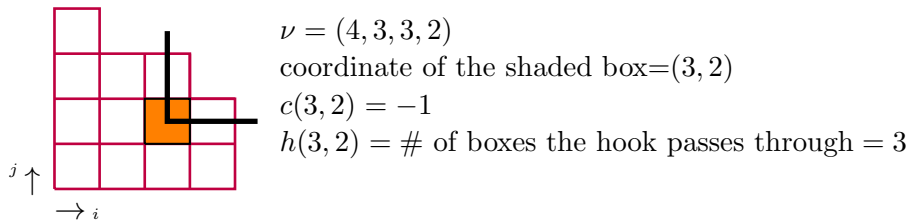
\begin{figure}\begin{center} $\begin{array}{c@{\hspace{1in}}c}
\multicolumn{1}{l}{\mbox{}} &
    \multicolumn{1}{l}{\mbox{}} \\ [-0.53cm]
{
\begin{pspicture}(0,0)(4,5)


\psline[unit=0.6cm,linewidth=1pt,linecolor=purple,fillstyle=none,fillcolor=white](0,0)(4,0)(4,2)(3,2)(3,3)(1,3)(1,4)(0,4)(0,0)
\psline[unit=0.6cm,linewidth=1pt,linecolor=purple,fillstyle=none,fillcolor=white](0,1)(4,1)
\psline[unit=0.6cm,linewidth=1pt,linecolor=purple,fillstyle=none,fillcolor=white](0,2)(4,2)
\psline[unit=0.6cm,linewidth=1pt,linecolor=purple,fillstyle=none,fillcolor=white](0,3)(1,3)
\psline[unit=0.6cm,linewidth=1pt,linecolor=purple,fillstyle=none,fillcolor=white](1,0)(1,3)
\psline[unit=0.6cm,linewidth=1pt,linecolor=purple,fillstyle=none,fillcolor=white](2,0)(2,3)
\psline[unit=0.6cm,linewidth=1pt,linecolor=purple,fillstyle=none,fillcolor=white](3,0)(3,3)
\psline[unit=0.6cm,fillstyle=solid,fillcolor=orange](2,1)(3,1)(3,2)(2,2)(2,1)
\psline[unit=0.6cm,linewidth=2pt,fillstyle=none,fillcolor=orange](2.5,3.5)(2.5,1.5)(4.5,1.5)

 \put(0,-0.4){$\rightarrow$} \put(-.4,0){$\uparrow$}
\put(-0.6,0.2){\tiny $j$} \put(0.5,-0.4){\tiny $i$} \put(3,2){\small
$\nu=(4,3,3,2)$} \put(3,1.5){\mbox{\small coordinate of the shaded
box=$(3,2)$}} \put(3,1){\small $c(3,2)=-1$} \put(3,0.5){\small
$h(3,2)=\mbox{\# of boxes the hook passes through}=3$}
\end{pspicture}
}
\\ [-0.5cm] \mbox{} & \mbox{}
\end{array}$\caption{The content and the hook length of a box in a Young diagram.} \label{contenthook}\end{center}
\end{figure}

Similarly for the Hopf link we can color the two component knots by
two different representations to obtain \bea
W_{\lambda\,\mu}=\langle \mbox{Tr}_{\lambda}U_1\,\mbox{Tr}_{\mu}
U_2\rangle \eea where $U_1$ and $U_2$ are the two holonomy matrices
around two component unknots. This can also be calculated exactly
 to obtain \bea
W_{\lambda\,\mu}=q_{1}^{\frac{\kappa(\mu)}{2}}\,s_{\lambda}(q_{1}^{-\rho})\,s_{\mu}(q_{1}^{-\rho-\lambda},Q\,q_{1}^{\rho})\prod_{(i,j)\in
\lambda}(1-Q\,q_{1}^{i-j}) .\eea Here $Q=q_{1}^{-N}$. We will recall
the geometry of D-branes for the unknot and Hopf link in section 4
and review how the open topological string amplitudes in the
presence of these branes reproduce the above knot and link
invariants, before extending it to more refined invariants.

In \cite{OV} the open topological string amplitudes were interpreted
as counting a certaion BPS partition function. This interpretation
is crucial for connecting it to link homologies as the Hilbert space
is naturally in the problem.  Moreover the gradation of the
homology, is nothing but the charges of BPS states in the physical
theory.  The geometry considered in \cite{OV} was as follows: We can
lift the type IIA geometry of the resolved conifold to M-theory.  In
this context the probe branes get mapped to M5 branes wrapping the
Lagrangian cycles and filling the non-compact $R^3$ spacetime.  The
open topological string simply compute the number of M2 branes
ending on the M5 branes.  The representation of the link invariant
encodes the geometry of the ending of the M2 brane on the M5 brane.
Moreover the coefficient of $q^s Q^J$ in the topological string
amplitudes, $N_{R,J,s}$, is determined by the number of such bound
states which wrap the $\mathbb{P}^1$ $J$ times and have spin $s$
under the SO(2) rotation of the spatial $\mathbb{R}^2\subset
\mathbb{R}^3$ \footnote{For a complete mathematical proof of the
integrality of $N_{R,J,s}$ see \cite{LP}.}.The precise structure of
the connection between open topological strings and BPS counting was
further elaborated in \cite{LMV}, to which we refer the interested
reader. For a single knot, for example, one finds that the free
energy $F={\rm log}(Z)$ as a function of $V$ defined above, is given
by
$$F(V)=-\sum_{R,n>0} f_R(q^n,Q^n) {{\rm Tr}_R V^n\over n}$$
where $f_R(q,Q)$ is completely determined by the BPS degeneracies
of the M2 brane, $N_{R',J,s}$, where $R'$ denotes the representation
the BPS state transforms in, $J$, is the charge of the brane
and $s$ is the spin.  Moreover the sign of $N$ is correlated with
its fermion number.

It was proposed in \cite{GSV} that there is a further charge one can
consider in labeling the BPS states of M2 branes ending on M5
branes:  The normal geometry to the M5 brane includes, in addition
to the spacetime $\mathbb{R}^3$, and the three normal directions
inside the CY, an extra $\mathbb{R}^2$ plane.  It was proposed there
that the extra $SO(2)$ rotation in this plane will provide an extra
gradation which could be viewed as a refinement of topological
strings and it was conjectured that this is related to link
homologies that we will review in the next section.  This gives
a refinement of $N_{R,J,s}\rightarrow N_{R,J,r,s}$.  In other words
for a given representation $R$ we have a triply graded structure
labeling the BPS states.


\section{\sc Link Homologies and Topological Strings}
\label{sec3}

Now, let us proceed to describing the properties of link homologies
suggested by the their relation to Hilbert spaces of BPS states.
We mostly follow notations of \cite{GSV,DGR}.

Let $L$ be an oriented link in ${\bf S}^3$
with $\ell$ components, $K_1, \ldots, K_{\ell}$.
We shall consider homological as well as polynomial invariants of $L$
whose components are colored by representartions $R_1, \ldots, R_{\ell}$
of the Lie algebra ${\bf g}$.
Although in this paper we shall consider only ${\bf g} = sl(N)$,
there is a natural generalization to other classical Lie algebras
of type $B$, $C$, and $D$.
In particular, there are obvious analogs of the structural properties
of $sl(N)$ knot homologies for $so(N)$ and $sp(N)$ homologies
(see \cite{GWalcher,RKhovanoviii} for some work in this direction).

Given a link colored by a collection of representations
$R_1,\ldots, R_{\ell}$ of $sl(N)$, we denote the corresponding polynomial
invariant by
\bea
\overline P_{sl(N);R_1, \ldots, R_{\ell}} (q)
\label{slnrrr}
\eea
Here and below, the ``bar'' means that \eqref{slnrrr} is the unnormalized invariant;
its normalized version $P_{sl(N);R_1, \ldots, R_{\ell}} (q)$ obtained
by dividing by the invariant of the unknot is written without a bar.
Since this ``reduced'' version depends on the choice of the ``preferred''
component of the link $L$, below we mainly consider a more natural,
unnormalized invariant \eqref{slnrrr}.
In the special case when every $R_a$, $a=1,\ldots,\ell$ is the
fundamental representation of $sl(N)$ we simply write
\bea
\overline P_N (q) \equiv \overline P_{sl(N);\tableau{1}, \cdots, \tableau{1}} (q)
\eea
The polynomial invariants \eqref{slnrrr} are related to
expectation values of Wilson loop operators
$W (L) = W_{R_1, \cdots, R_{\ell}} (L)$ in Chern-Simons theory.
For example, the polynomial $sl(N)$ invariant $P_N (q)$
is related to the expectation value of the Wilson loop operator
$W (L) = W_{\tableau{1}, \cdots, \tableau{1}} (L)$,
\bea
\bar P_N (L) = q^{-2N \lk (L)} \langle W (L) \rangle
\label{pnlinks}
\eea
where $\lk (L) = \sum_{a<b} \lk (K_a,K_b)$ is the total linking number of $L$.

Now, let us turn to the corresponding homological invariants.
Let $\mathcal{H}_{i,j}^{sl(N);R_1, \ldots, R_{\ell}} (L)$
be the doubly-graded homology theory whose graded Euler characteristic
is the polynomial invariant $\overline P_{sl(N);R_1, \ldots, R_{\ell}} (q)$,
\bea
\overline P_{sl(N);R_1, \ldots, R_{\ell}} (q) = \sum_{i,j \in \mathbb{Z}} (-1)^j q^i
\dim \mathcal{H}_{i,j}^{sl(N);R_1, \ldots, R_{\ell}} (L)
\label{pslncategor}
\eea
The graded Poincar\'e polynomial,
\bea
\overline{\mathcal{P}}_{sl(N);R_1, \ldots, R_{\ell}} (q,t) :=
\sum_{i,j \in \mathbb{Z}} q^i t^j
\dim \mathcal{H}_{i,j}^{sl(N);R_1, \ldots, R_{\ell}} (L)
\label{superpsln}
\eea
is, by definition, a polynomial in $q^{\pm 1}$ and $t^{\pm 1}$
with integer non-negative coefficients.
Clearly, evaluating \eqref{superpsln} at $t=-1$ gives \eqref{pslncategor}.

When $R_a = \tableau{1}$ for all $a=1, \ldots, \ell$,
the homology $\mathcal{H}_{i,j}^{sl(N);R_1, \ldots, R_{\ell}} (L)$
is the Khovanov-Rozansky homology, $\overline{HKR}_{i,j}^N (L)$, and
\bea
& \overline{KhR}_N (q,t) \equiv \overline{\mathcal{P}}_{sl(N);\tableau{1}, \cdots, \tableau{1}} (q,t) \cr
& ~~~~~~~~~~~~~~~~~~~~~~~~~~~~ = \sum_{i,j \in \mathbb{Z}} q^i t^j \dim \overline{HKR}_{i,j}^N (L)
\eea
is its graded Poincar\'e polynomial.

The physical interpretation of homological link invariants via
Hilbert spaces of BPS states leads to certain predictions regarding
the behavior of link homologies with rank $N$.
In particular, the total dimension of
$\mathcal{H}_{*,*}^{sl(N);R_1, \ldots, R_{\ell}} (L)$ grows as
\bea
\dim \mathcal{H}_{*,*}^{sl(N);R_1, \ldots, R_{\ell}} (L) \sim N^{d}
\quad,\quad N \to \infty
\label{hdimgrowth}
\eea
where
\bea
d = \sum_{i=1}^{\ell} \dim R_i
\eea
More specifically, a general form of the conjecture in \cite{GSV} states:

{\bf Conjecture:} {\it There exists a ``superpolynomial''
$\overline{\mathcal{P}}_{R_1, \ldots, R_{\ell}} ({\bf a},q,t)$,
a rational function\footnote{This definition differs slightly from
the ones introduced in \cite{DGR}, where it is the numerator of the
rational function $\overline{\mathcal{P}}_{R_1, \ldots, R_{\ell}} ({\bf a},q,t)$
which was called the superpolynomial. Since in general one has a very good
control of the denominators, the two definitions are clearly related.}
in three variables ${\bf a}$, $q$, and $t$, such that}
\bea
\overline{\mathcal{P}}_{sl(N);R_1, \ldots, R_{\ell}} (q,t) =
\overline{\mathcal{P}}_{R_1, \ldots, R_{\ell}} ({\bf a}=q^N,q,t)
\label{superp}
\eea
{\it for sufficiently large $N$.}

The coefficients of the superpolynomial, say,
in the case of the fundamental representation:
\bea
\overline{\mathcal{P}}_{N} ({\bf a},q,t)
= {1 \over (q-q^{-1})^{\ell}} \sum_{J,s,r} {\bf a}^J q^s t^r D_{J,s,r}
\eea
encode the dimensions of the Hilbert space of states, related to BPS states,
\bea
D_{J,s,r} := (-1)^F \dim \mathcal{H}_{BPS}^{F,J,s,r}
\eea
graded by the fermion number $F$, the membrane charge $J$,
and the $U(1)_L \times U(1)_R$ quantum numbers $s$ and $r$.  However,
note that the $D_{J,s,r}$ is not the same as $N_{J,s,r}$:  $N_{J,s,r}$
encodes the integral structure in the Free energy, whereas
$D_{J,s,r}$ is the exponentiated version of it.  It is not
difficult to see that the integrality of $N_{J,s,r}$ guarantees that
of $D_{J,s,r}$ (as in closed string case where
the integrality of GV invariants implies integrality of the DT
invariants).  This in particular
explains that the Hilbert space
structure of BPS states captured by $N_{J,s,r}$ will indeed encode
the Hilbert space structure for $D_{J,s,r}$ and thus its integrality.
However, it is not completely obvious from the physical picture
why \eqref{superp} is a {\it finite} polynomial, for any given $N$,
as has been conjectured.

The conjecture \eqref{superp} can be refined even further. Indeed, the large $N$ growth
described in \eqref{hdimgrowth} and \eqref{superp} is characterized by the contribution
of individual link components,
\bea
\oplus_{a=1}^{\ell} \mathcal{H}_{*,*}^{sl(N);R_a} (K_a)
\eea
Often, it is convenient to remove this contribution and consider
only the ``connected'' part of the polynomial (resp. homological)
link invariant. For example, in the simplest case when all
components of the link $L$ carry the fundamental representation,
the corresponding $sl(N)$ invariant $\bar P_N (L)$ or, equivalently,
the Wilson loop correlation function \eqref{pnlinks} can be written
in terms of the integer BPS invariants $N_{(\tableau{1}, \cdots, \tableau{1}),Q,s}$
as
\bea
\langle W (L) \rangle^{(c)} = (q^{-1}-q)^{\ell-2}
\sum_{J,s} N_{(\tableau{1}, \cdots, \tableau{1}),J,s} q^{NJ+s}
\label{wcvian}
\eea
where $\langle W (L) \rangle^{(c)}$ is the {\it connected} correlation function.
Thus, for a two-component link, we have
\bea
\langle W (L) \rangle^{(c)} =
\langle W (L) \rangle - \langle W (K_1) \rangle \langle W (K_2) \rangle
\label{wconnected}
\eea
and
\bea
\bar P_N (L) = q^{-2N \lk (L)} \Big[ \bar P_N (K_1) \bar P_N (K_2)
+ \sum_{J,s} N_{(\tableau{1}, \tableau{1}),J,s} q^{NJ+s} \Big]
\label{pntwocomp}
\eea
where $\bar P_N (K_1)$ and $\bar P_N (K_2)$ denote
the unnormalized $sl(N)$ polynomials of the individual link components.

Similarly, the homological $sl(N)$ invariant of a two-component link $L$
can be written as a sum of connected and disconnected terms \cite{GSV}:
\bea
\overline{KhR}_N (L) = q^{-2N \lk (L)}
\Big[ t^{\a} \overline{KhR}_N (K_1) \overline{KhR}_N (K_2)
+ {1 \over q-q^{-1}} \sum_{J,s,r \in \mathbb{Z}} D_{J,s,r} q^{NJ+s} t^r
\Big]
\label{khtwocompviad}
\eea
where integer invariants $D_{J,s,r} (L)$ are related to
the dimensions of the Hilbert space of BPS states, $N_{J,s,r}$
and $\a$ is a simple invariant of $L$. 
At $t=-1$ this expression specializes to (\ref{pntwocomp}).

\subsection{\sc Hopf link: the fundamental representation}

The Hopf link, $L=2_1^2$ consists of two components, $K_1 \cong K_2 \cong {\rm unknot}$,
which are linked with the linking number $\lk (K_1, K_2) = -1$.
The $sl(2)$ homological invariant for the Hopf link is:
\bea
\overline{KhR}_{2} (2_1^2) = 1 + q^2 + q^4 t^2 + q^6 t^2
\label{khhopfsltwo}\eea
It can be written in the form  \eqref{khtwocompviad}
with the following non-zero invariants
\bea
& D_{0,-1,0} = 1, \quad D_{0,1,2} = -1, \quad \cr
& D_{-2,-1,0} = -1, \quad D_{-2,1,2} = 1
\label{guesshopf}
\eea
This gives the ``superpolynomial'' for the Hopf link,
\bea
\overline{\mathcal P} (2_1^2) =
{1 \over (q-q^{-1})^2} \Big[
\left( q^{-2} - 1 + q^2 t^2 \right)
 + {\bf a}^2 \left( 1 - q^2 t^2 - q^{-2} - t^2 \right)
 + {\bf a}^4 t^2 \Big]
\label{khhopfrefined}
\eea
which after specializing to ${\bf a}=q^N$ gives the graded Poincar\'e
polynomial of the $sl(N)$ link homology:
\bea
\overline{KhR}_{N} (2_1^2) =
q^{N-1} \Big( {q^N - q^{-N} \over q - q^{-1}} \Big)
+ q^{2N} \Big( {q^N - q^{-N} \over q - q^{-1}} \Big)^2 t^2
- q^{N+1} \Big( {q^N - q^{-N} \over q - q^{-1}} \Big) t^2
\label{khhopfn}
\eea
Notice, that at $t=-1$ this expression reduces
to the correct formula for the $sl(N)$ polynomial
invariant of the Hopf link,
\bea
\overline P_N (2_1^2) = 1 - q^{2N}
+ q^{2N} \Big( {q^N - q^{-N} \over q - q^{-1}} \Big)^2
\label{pnhopf}
\eea
The result (\ref{khhopfn}) agrees with the direct computation
of Khovanov-Rozansky homology for small values of $N$:
\bea\nn
\overline{KhR}_{3}(2_1^2)&=& 1 + q^2 + q^4 + q^4 t^2
+ 2 q^6 t^2 + 2 q^8 t^2 + q^{10} t^2 \\\nn
\overline{KhR}_{4}(2_1^2)&=& 1 + q^2 + q^4 + q^4 t^2
+ q^6 + 2 q^6 t^2 + 3 q^8 t^2 + 3 q^{10} t^2 + 2 q^{12} t^2
+ q^{14} t^2 \\\nn
\overline{KhR}_{5}(2_1^2)&=& 1 + q^2 + q^4 + q^4 t^2
+ q^6 + 2 q^6 t^2 + q^8 + 3 q^8 t^2 + 4 q^{10} t^2 + 4 q^{12} t^2 \\
&&+ 3 q^{14} t^2 + 2 q^{16} t^2 + q^{18} t^2
\label{hopfsmalln}
\eea

\section{\sc Refined topological vertex}
\label{sec4}

In this section we will breifly explain the combinatorial interpretation of
the refined vertex in terms of 3D partitions; more details can be found in \cite{IKV}.

\begin{figure}
$\begin{array}{c@{\hspace{1in}}c} \multicolumn{1}{l}{\mbox{}} &
    \multicolumn{1}{l}{\mbox{}} \\ [-0.53cm]
 \includegraphics[width=2.5in]{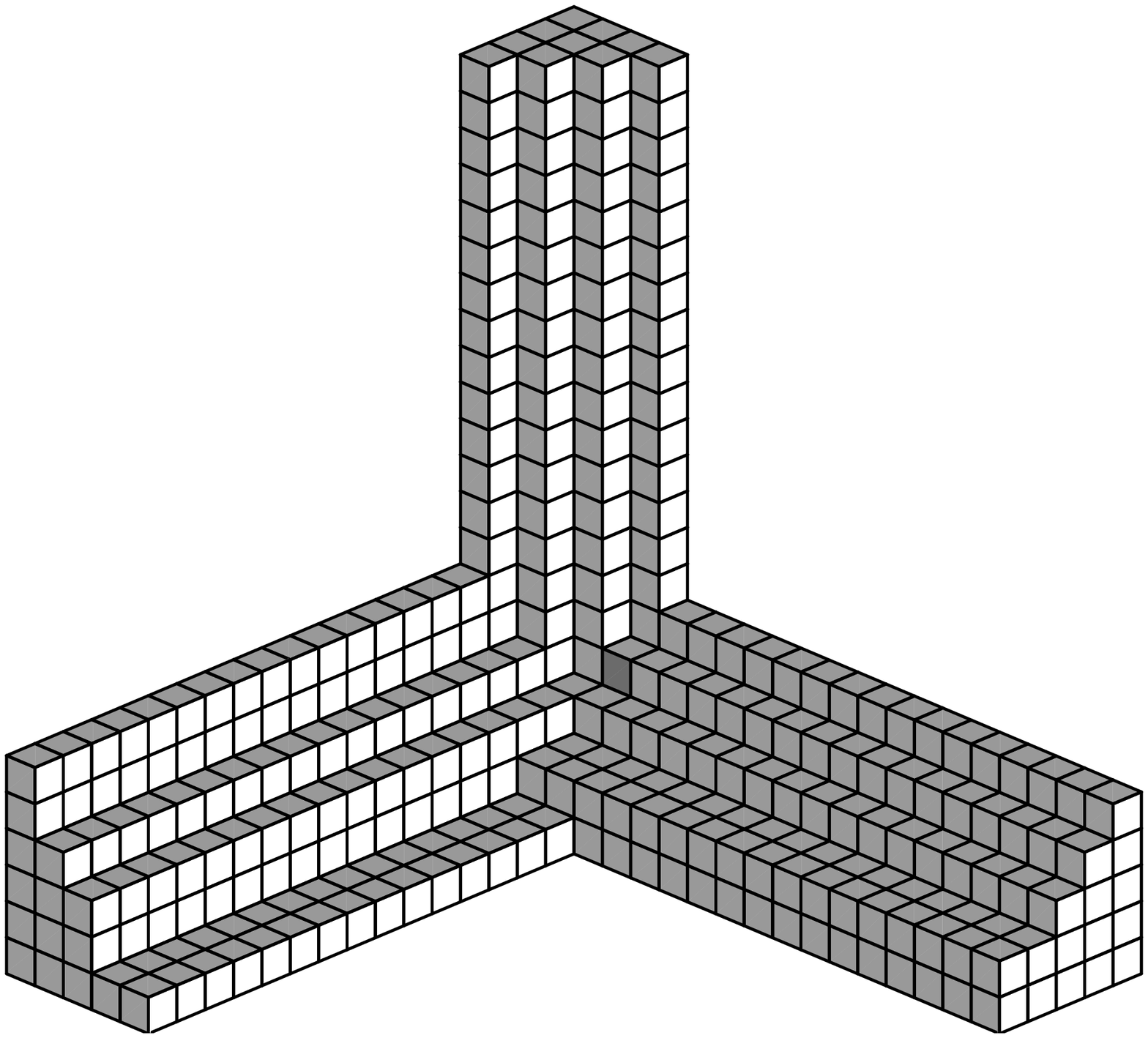} &
\includegraphics[width=2.5in]{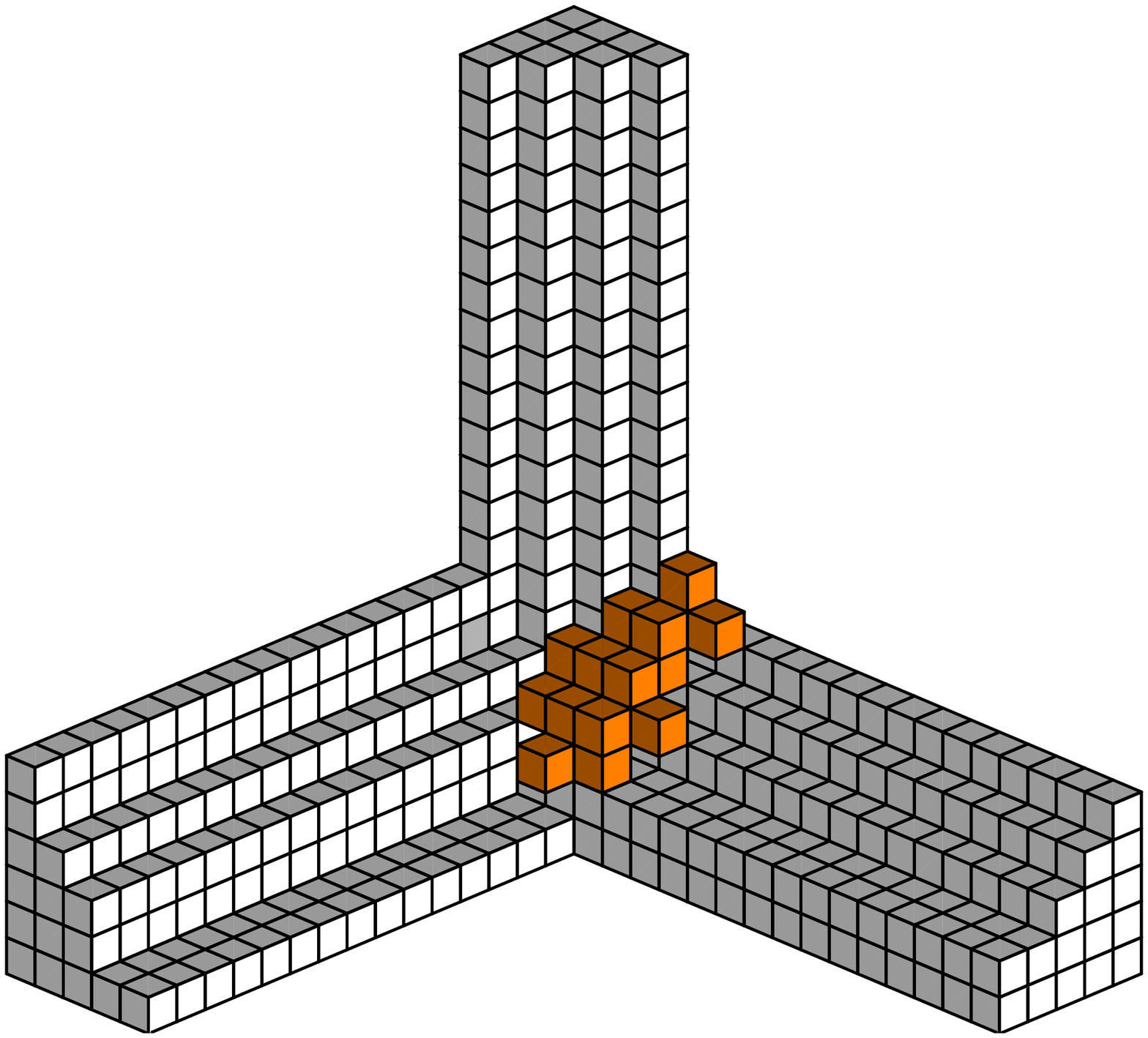}\\
\\ [0.4cm] \mbox{(a)} & \mbox{(b)}
\end{array}$
\caption{(a) $\pi_{\bullet}(\lambda,\mu,\nu)$ for
$\lambda=(6,4,3,1,1),\mu=(5,4,3,2,2),\nu=(4,3,2,1)$.  (b) An example
of $\pi(\lambda,\mu,\nu)$ } \label{vertex}
\end{figure}

Recall that the generating function of the 3D partitions is given by
the MacMahon function, \bea M(q)&=&\sum_{n\geq
0}C_{n}q^{n}=\prod_{k=1}^{\infty}(1-q^{n})^{-n}\,,\\\nn C_{n}&=&
\mbox{\# of 3D partitions with $n$ boxes} \eea The topological
vertex $C_{\lambda\,\mu\,\nu}(q)$ \cite{AKMV} \bea
C_{\lambda\,\mu\,\nu}(q)=q^{\frac{\kappa(\mu)}{2}}\,s_{\nu^{t}}(q^{-\rho})\,\sum_{\eta}s_{\lambda^{t}/\eta}(q^{-\rho-\nu})\,s_{\mu/\eta}(q^{-\rho-\nu^{t}})\,,
\eea has the following combinatorial interpretation \cite{ORV} \bea
M(q)C_{\lambda\,\mu\,\nu}(q)=f_{\lambda\,\mu\,\nu}(q)\,\sum_{\pi(\lambda,\mu,\nu)}q^{|\pi(\lambda,\mu,\nu)|-|\pi_{\bullet}(\lambda,\mu,\nu)|}\,.
\eea Where $\pi(\lambda,\mu,\nu)$ is a 3D partition such that along
the three axis which asymptotically approaches the three 2D
partitions $\lambda,\mu$ and $\nu$. $|\pi|$ is number of boxes
(volume) of the 3D partition $\pi$ and $\pi_{\bullet}$ is the 3D
partition with the least number of boxes satisfying the same
boundary condition\footnote{Since even the partition with the least
number of boxes has infinite number of boxes we need to regularize
this by putting it in an $N\times N\times N$ box as discussed in
\cite{ORV}}. \figref{vertex}(a) shows the $\pi_{\bullet}$ for
$\lambda=(6,4,3,1,1),\mu=(5,4,3,2,2)$ and $\nu=(4,3,2,1)$.
\figref{vertex}(b) shows an example of the partition
$\pi(\lambda,\mu,\nu)$ for $\lambda,\mu,\nu$ same as in the
\figref{vertex}(a). $f_{\lambda\,\mu\,\nu}(q)$ is the framing factor
which appears because of the change from perpendicular slicing of
the 3D partition to diagonal slicing of the 3D partition \cite{ORV}.

The refined topological vertex \cite{IKV} \bea
C_{\lambda\,\mu\,\nu}(q_{1},q_{2})&=&\Big(\frac{q_{1}}{q_{2}}\Big)^{\frac{||\mu||^2-|\mu|}{2}}\,q_{2}^{\frac{\kappa(\mu)}{2}}\,
q_{2}^{\frac{||\nu||^2}{2}}\,\widetilde{Z}_{\nu}(q_{1},q_{2})\\\nn
&&\times
\sum_{\eta}\,\Big(\frac{q_{2}}{q_{1}}\Big)^{\frac{|\eta|+|\lambda|-|\mu|}{2}}\,
s_{\lambda^{t}/\eta}(q_{1}^{-\rho}\,q_{2}^{-\nu})\,s_{\mu/\eta}(q_{2}^{-\rho}\,q_{1}^{-\nu^{t}})\,
\label{vv} \eea also has a similar combinatorial interpretation in
terms of 3D partitions which we will explain now. Recall that the
diagonal slices of a 3D partition, $\pi$, are 2D partitions which
interlace which each other. These are the 2D partitions living on
the planes $x-y=a$ where $a\in \mathbb{Z}$. We will denote these 2D
partitions by $\pi_{a}$. For the usual vertex the a-th slice is
weighted with $q^{|\pi_{a}|}$ where $|\pi_{a}|$ is the number of
boxes cut by the slice (the number of boxes in the 2D partition
$\pi_{a}$). The 3D partition is then weighted by\bea \prod_{a\in
\mathbb{Z}}q^{|\pi_{a}|}=q^{\sum_{a\in
\mathbb{Z}}|\pi_{a}|}=q^{\mbox{$\#$ of boxes in the $\pi$}} \eea In
the case of the refined vertex the 3D partition is weighted in a
different manner. Given a 3D partition $\pi$ and its diagonal slices
$\pi_{a}$ we weigh the slices for $a<0$ with parameter $q$ and the
slices with $a\geq 0$ with parameter $t$ so that the measure
associated with $\pi$ is given by \bea
\Big(\prod_{a<0}q_{2}^{|\pi_{a}|}\Big)\,\Big(\prod_{a\geq
0}q_{1}^{|\pi_{a}|}\Big)=q_{2}^{\sum_{i=1}^{\infty}|\pi(-i)|}\,q_{1}^{\sum_{j=1}^{\infty}|\pi(j-1)|}\,.
\eea The generating function for this counting is a generalization
of the MacMahon function and is given by \bea M(q_{1},q_{2}):=
\sum_{\pi}q_{2}^{\sum_{i=1}^{\infty}|\pi(-i)|}\,q_{1}^{\sum_{j=1}^{\infty}|\pi(j-1)|}=\prod_{i,j=1}^{\infty}(1-q_{1}^{j}q_{2}^{i-1})^{-1}\,.
\eea We can think of this assignment of $q_{1}$ and $q_{2}$ to the
slices in the following way. If we start from large positive $a$ and
moves toward the slice passing through the origin then every time we
move the slice towards the left we count it with $q_{1}$ and every
time we move the slice up (which happens when we go from $a=i$ to
$a=i-1$, $i=0,1,2\cdots$) we count it with $q_{2}$.

\begin{figure}\begin{center}
$\begin{array}{c@{\hspace{1in}}c} \multicolumn{1}{l}{\mbox{}} &
    \multicolumn{1}{l}{\mbox{}} \\ [-0.53cm]
{
\begin{pspicture}(9,0)(4,5)
\psline[unit=0.75cm,linecolor=red,linestyle=dashed](5,0)(6.2,1.2)
\psline[unit=0.75cm,linecolor=red, linestyle=dashed](4,0)(6.2,2.2)
\psline[unit=0.75cm,linecolor=red,linestyle=dashed](3,0)(6.2,3.2)

\psline[unit=0.75cm,linecolor=blue](3,1)(6.2,4.2)
\psline[unit=0.75cm,linecolor=red,linestyle=dashed](2,1)(6.2,5.2)
\psline[unit=0.75cm,linecolor=blue](2,2)(6.2,6.2)
\psline[unit=0.75cm,linecolor=blue](2,3)(5.2,6.2)
\psline[unit=0.75cm,linecolor=red,linestyle=dashed](1,3)(4.2,6.2)
\psline[unit=0.75cm,linecolor=blue](1,4)(3.2,6.2)
\psline[unit=0.75cm,linecolor=red,linestyle=dashed](0,4)(2.2,6.2)
\psline[unit=0.75cm,linecolor=blue](0,5)(1.2,6.2)
\psline[unit=0.75cm,linecolor=red,linestyle=dashed](6,0)(6.2,0.2)
\psline[unit=0.75cm,linecolor=blue](0,6)(.2,6.2)
\psline[unit=0.75cm,linecolor=myorange](0,0)(0,7)
\psline[unit=0.75cm,linecolor=myorange](0,0)(7,0)

\psline[unit=0.75cm,linewidth=2pt,linecolor=purple,fillstyle=none,fillcolor=white](0,0)(3,0)(3,1)(2,1)(2,3)(1,3)(1,4)(0,4)(0,0)

\put(0.3,0.2){\small{$5$}}
\put(0.3,1){\small{$3$}}
\put(0.3,1.7){\small{$2$}}
\put(0.27,2.5){\small{$1$}}
\put(1,0.2){\small{$4$}}
\put(1.7,0.2){\small{$2$}}
\put(1,1){\small{$3$}}
\put(1,1.7){\small{$1$}}

\put(5,4){{\large$\pi\rightarrow
\,\prod_{a\in\mathbb{Z}}q_{a}^{|\pi(a)|}=q_{2}^{\sum_{i=1}^{\infty}|\pi(\nu^{t}_{i}-i)|}\,q_{1}^{\sum_{j=1}^{\infty}|\pi(-\nu_{j}+j-1)|}$}}
\put(5,3){{\large
$=q_{2}^{\pi(2)+\pi(0)+\pi(-1)+\pi(-3)+\sum_{i=5}^{\infty}\pi(-i)}$}}
\put(5.2,2){$\times ${\large $\,\,
q_{1}^{\pi(-4)+\pi(-2)+\pi(1)+\sum_{j=4}^{\infty}\pi(j-1)}=q_{2}^{15}\,q_{1}^{6}$}}

\put(5,1){$q_{2}=\mbox{blue}\,(\mbox{solid
line}),\,q_{1}=\mbox{red}\,(\mbox{dashed line})$}

\put(5,0.5){$\nu=(4,3,1)$}

\psgrid[unit=0.75cm, subgriddiv=1,
gridcolor=myorange, %
gridlabelcolor=white]%
(0,0)(6,6)
\end{pspicture}
}
\\ [-0.5cm] \mbox{} & \mbox{}
\end{array}$
\caption{Slices of the 3$D$ partitions are counted with parameters
$q_{1}$ and $q_{2}$ depending on the shape of $\nu$.}
\label{ex}\end{center}
\end{figure}
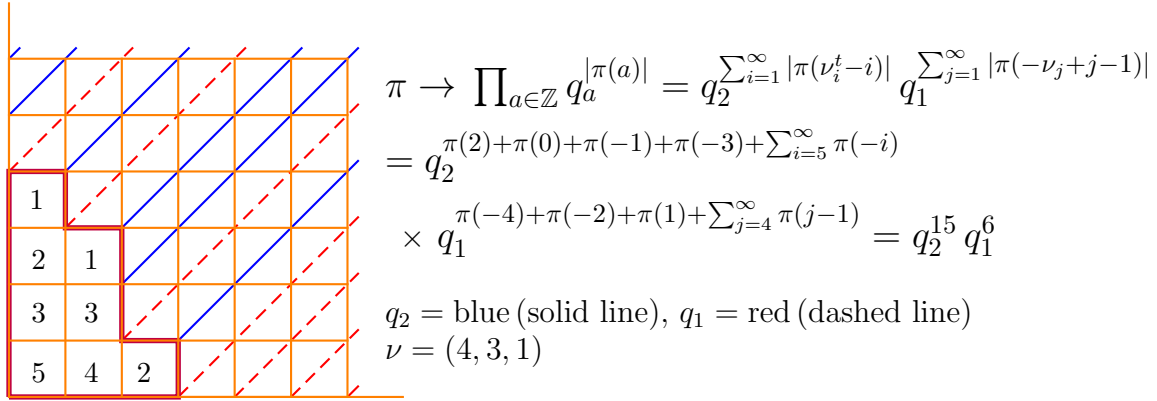

Since we are slicing the skew 3D partitions with planes $x-y=a$ we
naturally have a preferred direction given by the z-axis. We take
the 2D-partition along the z-axis to be $\nu$. The case we discussed
above, obtaining the refined MacMahon function, had $\nu=\emptyset$.
For non-trivial $\nu$ the assignment of $q_{2}$ and $q_{1}$ to various
slices is different and depends on the shape of $\nu$. As we go from
$+\infty$ to $-\infty$ the slices are counted with $q_{1}$ if we go
towards the left and is counted with $q_{2}$ if we move up. An example
is shown in \figref{ex}.

After taking into account the framing and the fact that the slices
relevant for the topological vertex are not the perpendicular slices
\cite{ORV} the generating function is given by \bea\nn
G_{\lambda\,\mu\,\nu}(q_{1},q_{2})=M(q_{1},q_{2})\times
C_{\lambda\,\mu\,\nu}(q_{1},q_{2}) \eea where
$C_{\lambda\,\mu\,\nu}(q_{1},q_{2})$ is the refined topological
vertex \bea\nn
C_{\lambda\,\mu\,\nu}(q_{1},q_{2})&=&\Big(\frac{q_{2}}{q_{1}}\Big)^{\frac{||\nu||^2-||\nu||^2}{2}}\,q_{2}^{\frac{\kappa(\mu)}{2}}\,
P_{\nu^{t}}(q_{1}^{-\rho};q_{2},q_{1})\,
\sum_{\eta}\Big(\frac{q_{2}}{q_{1}}\Big)^{\frac{|\eta|+|\lambda|-|\mu|}{2}}s_{\lambda^{t}/\eta}(q_{1}^{-\rho}q_{2}^{-\nu})
s_{\mu/\eta}(q_{1}^{-\nu^{t}}q_{2}^{-\rho}) \eea

In the above expression $P_{\nu}({\bf x};q_{2},q_{1})$ is the
Macdonald function such that \bea
P_{\nu^{t}}(q_{1}^{-\rho};q_{2},q_1)&=&q_{1}^{\frac{||\nu||^{2}}{2}}\,\widetilde{Z}_{\nu}(q_1,q_2)\,,\\\nn
\widetilde{Z}_{\nu}(q_1,q_2)&=&\prod_{(i,j)\in
\,\nu}\Big(1-q_{1}^{a(i,j)+1}\,q_{2}^{\ell(i,j)}\Big)^{-1}\,,\,\,\,\,
a(i,j)=\nu^{t}_{j}-i\,,\,\,\,\ell(i,j)=\nu_{i}-j\,. \eea

\subsection{\sc open topological string amplitudes}

In this section we will discuss the open string partition function
obtained from the topological vertex and their relation with
polynomial Hopf link invariants. Recall that the usual topological
vertex is given by \cite{AKMV,ORV} \bea
C_{\lambda\,\mu\,\nu}(q_1)=\,q_1^{\frac{\kappa(\mu)}{2}}\,s_{\nu^{t}}(q_1^{-\rho})\sum_{\eta}\,s_{\lambda^{t}/\eta}(q_1^{-\rho-\nu})\,s_{\mu/\eta}(q_1^{-\rho-\nu^{t}})
\eea Although written in terms of the Schur and skew-Schur functions
in the above equation it can be rewritten in terms of $sl(N)$ Hopf
link invariants for large $N$ \cite{AKMV}, \bea {\cal
W}_{\lambda\,\mu}(q_1)=q_1^{-\frac{\kappa(\mu)}{2}}\,C_{\lambda^{t}\,\mu\,\varnothing}(q_1)=\,\left\{
                      \begin{array}{lll}
                        \sum_{\eta}s_{\lambda/\eta}(q_1^{-\rho})\,s_{\mu/\eta}(q_1^{-\rho})\,&\\
                         q_1^{-\frac{\kappa(\mu)}{2}}\,s_{\lambda}(q_1^{-\rho})\,s_{\mu^{t}}(q_1^{-\rho-\lambda}) &\\
                        q_1^{-\frac{\kappa(\mu)+\kappa(\lambda)}{2}}\,s_{\mu^{t}}(q_1^{-\rho})\,s_{\lambda^{t}}(q_1^{-\rho-\mu^{t}})\,.&
                      \end{array}
                    \right.
 \label{hopfvertex}\eea The above three expressions are equivalent because of cyclic symmetry of the topological vertex. Next, we
will show that $sl(N)$ Hopf link invariants can be related
to the open string partition function calculated using the
topological vertex. Eq(\ref{hopfvertex}) will guide us in
formulating the precise relation between the $sl(N)$ Hopf link
invariant and the open string partition function.

\subsubsection{\sc Hopf link}

\begin{figure}\begin{center} $\begin{array}{c@{\hspace{1in}}c}
\multicolumn{1}{l}{\mbox{}} &
    \multicolumn{1}{l}{\mbox{}} \\ [-0.53cm]
{

\begin{pspicture}(4,-1)(6,3)
\psline[unit=0.75cm,linecolor=black,linewidth=2pt](1.5,1.5)(5.5,1.5)
\psline[unit=0.75
cm,linecolor=black,linewidth=2pt](3.5,-0.3)(3.5,1.3)
\psline[unit=0.75cm,linecolor=black,linewidth=2pt](3.5,1.7)(3.5,3.5)
\psline[unit=0.75cm,linecolor=blue,linewidth=2pt,linestyle=dashed](2.5,1.5)(3.5,2.5)

\psline[unit=0.75cm,linecolor=violet]{->}(7.5,2)(9,2.5)
\psline[unit=0.75cm,linecolor=violet]{->}(7.5,1)(9,0.5)

\pscurve[unit=0.75cm,linecolor=violet]{<->}(14,0.5)(14.5,1.5)(14,2.5)

\put(11,1){\small{Flop transition}}

\psline[unit=0.75cm,linecolor=orange](12,3)(13,4)
\psline[unit=0.75cm,linecolor=orange](11,3)(12,3)
\psline[unit=0.75cm,linecolor=orange](12,2)(12,3)
\psline[unit=0.75cm,linecolor=orange](13,4)(13,5)
\psline[unit=0.75cm,linecolor=orange](13,4)(14,4)

\psline[unit=0.75cm,linecolor=orange](12,0)(13,-1)
\psline[unit=0.75cm,linecolor=orange](11,0)(12,0)
\psline[unit=0.75cm,linecolor=orange](12,0)(12,1)
\psline[unit=0.75cm,linecolor=orange](13,-1)(14,-1)
\psline[unit=0.75cm,linecolor=orange](13,-1)(13,-2)

\psline[unit=0.75cm,linecolor=purple,linestyle=dashed](2,1.45)(1,2.45)
\psline[unit=0.75cm,linecolor=purple,linestyle=dashed](3.5,0.3)(4.7,-0.2)
\psline[unit=0.75cm,linecolor=purple,linestyle=dashed](12,2.5)(13.1,2)
\psline[unit=0.75cm,linecolor=purple,linestyle=dashed](11.5,3)(10.8,4)
\psline[unit=0.75cm,linecolor=purple,linestyle=dashed](13,-1.5)(14,-2)
\psline[unit=0.75cm,linecolor=purple,linestyle=dashed](11.5,0)(10.8,1)
\put(0.8,1.2){$\lambda$} \put(2.8,-0.3){$\mu$}
\put(8.15,2.3){$\lambda$} \put(9.2,1.4){$\mu$}
\put(8.15,0){$\lambda$} \put(9.9,-1.6){$\mu$}
\psline[unit=0.75cm,linecolor=red,linestyle=solid,linewidth=2pt](12.4,-0.7)(12.7,-0.4)
\psline[unit=0.75cm,linecolor=red,linestyle=solid,linewidth=2pt](12.6,3.3)(12.3,3.6)
\put(11,2.8){$(a)$} \put(11,-1){$(b)$}
\end{pspicture}}
\\ [0.3cm] \mbox{} & \mbox{}
\end{array}$
\caption{Two different resolutions of the conifold related to each
other by flop transition. The normalized partition function of the
geometry (b) gives homological $sl(N)$ invariants of the Hopf link
decorated by representations $(R_1, R_2)$. The red mark indicates
the choice of the preferred direction for the refined vertex.}
\label{flop}\end{center}
\end{figure}
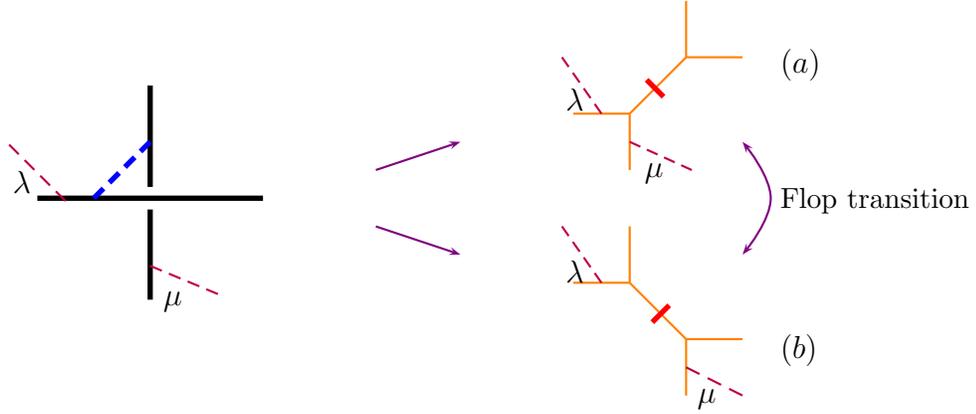

As we discussed in section 2, after geometric transition, the Hopf
link is represented by a pair of toric Lagrangian branes in the
geometry ${\cal O}(-1)\oplus {\cal O}(-1)\mapsto \mathbb{P}^{1}$.
Furthermore, as we also discussed earlier, there are two possible
resolutions of the singular conifold, both given by ${\cal
O}(-1)\oplus{\cal O}(-1)\mapsto \mathbb{P}^{1}$, related to each
other by a flop transition as shown in \figref{flop}. We will
determine the open string partition function for both these
configurations.

The open string partition function for the configuration shown in
\figref{flop}(a) is given by \bea Z^{\bf
I}(q_1,Q,V_{1},V_{2})=\sum_{\lambda\,,\,\mu}Z^{\bf
I}_{\lambda\,\,\mu}(q_1,Q)\,\mbox{Tr}_{\lambda}V_{1}\,\mbox{Tr}_{\mu}V_{2}\,,
\eea where $V_{1}$ and $V_{2}$ are the two holonomy matrices
associated with the two unknot components of the Hopf link and \bea
Z^{\,\,\bf
I}_{\lambda\,\mu}(q_1,Q)&=&\sum_{\nu}(-Q)^{|\nu|}\,C_{\lambda\,\mu\,\nu}(q_1)\,C_{\varnothing\,\varnothing\,\nu^{t}}(q_1)\\\nn
&=&\,s_{\lambda^{t}}(q_1^{-\rho})\,s_{\mu^{t}}(q_1^{-\rho-\lambda},Q\,q_1^{\rho})
\prod_{i,j=1}^{\infty}(1-Q\,q_1^{i+j-1-\lambda^{t}_{j}})\,. \eea We
normalize the above open string partition function by dividing with
the closed string partition function to obtain, \bea\nn
\widehat{Z}^{\,\,\bf I}_{\lambda\,\mu}(q_1,Q):=\frac{Z^{\,\,\bf
I}_{\lambda\,\mu}(q_1,Q)}{Z^{\,\,\bf
I}_{\varnothing\,\varnothing}(q_1,Q)}&=& \,
\,s_{\lambda^{t}}(q_1^{-\rho})\,s_{\mu^{t}}(q_1^{-\rho-\lambda},Q\,q_1^{\rho})\prod_{(i,j)\in
\lambda}(1-Q\,q_1^{j-i})\\\nn \eea In the limit $Q\mapsto 0$ we get
\bea \widehat{Z}^{\,\,\bf I}_{\lambda\,\mu}(q_1,Q=0)=\,
C_{\lambda\,\mu\,\varnothing}=q_1^{\frac{\kappa(\mu)}{2}}\,{\cal
W}_{\lambda^{t}\,\mu}(q_1)\,.\eea The right-hand side is the large
$N$ limit of the $sl(N)$ Hopf link invariant. The above equation
suggests the following relation between the open string partition
function and the $sl(N)$ Hopf link invariant, \bea
W_{\lambda\,\mu}(q_1,N)=q_1^{-\frac{\kappa(\mu)}{2}}\,\widehat{Z}^{\bf
I}_{\lambda^t\,\mu}(q_1,Q)\,,\,\,\,\,Q=q_1^N\,. \eea For
$(\lambda,\mu)=(\tableau{1}\,,\tableau{1})$ we get \bea
W_{\tableau{1}\,\tableau{1}}(q_1,N)=\widehat{Z}^{\,\,\bf
I}_{\tableau{1}\,\tableau{1}}(q_1,Q)&=&\,s_{\tableau{1}}(q_1^{-\rho-\tableau{1}},Q\,q_1^{\rho})\,q_1^{\frac{1}{2}}
\frac{1-Q}{1-q_1}\\\nn
&=&\Big(q_1^{-\frac{1}{2}}+\frac{q_1^{\frac{3}{2}}}{1-q_1}-Q\,\frac{q_1^{\frac{1}{2}}}{1-q_1}\Big)\,q_1^{\frac{1}{2}}\,\frac{1-Q}{1-q_1}\\\nn
&=&\,\frac{1-q_1+q_1^{2}}{(1-q_1)^{2}}-Q\,\frac{1+q_1^{2}}{(1-q_1)^{2}}+Q^{2}\,\frac{q_1}{(1-q_1)^{2}}\,.
\eea

{\bf Flop Transition:}\\
The other possibility for the geometry after transition is as shown
in \figref{flop}(b). In this case the partition function is given by
\bea Z^{\,\,\bf
II}_{\lambda\,\mu}(q_1,\widehat{Q})&=&\sum_{\nu}(-\widehat{Q})^{|\nu|}\,C_{\varnothing\,\mu\,\nu}(q_1)\,C_{\lambda\,\varnothing\,\nu^{t}}(q_1)\,\\\nn
&=&\,q_1^{\frac{\kappa(\mu)}{2}}\,\sum_{\nu}(-\widehat{Q})^{|\nu|}\,s_{\nu}(q_1^{-\rho})\,s_{\nu^{t}}(q_1^{-\rho})\,s_{\lambda^{t}}(q_1^{-\rho-\nu^{t}})
\,s_{\mu}(q_1^{-\rho-\nu^{t}})\, \eea For
$(\lambda,\mu)=(\tableau{1},\tableau{1})$ we get \bea
\widehat{Z}^{\,\,\bf
II}_{\tableau{1}\,\tableau{1}}(q_1,\widehat{Q})&=&\frac{Z^{\,\,\bf
II}_{\tableau{1}\,\tableau{1}}(q_1,\widehat{Q})}{Z^{\,\,\bf
II}_{\varnothing\,\varnothing}(q_1,\widehat{Q})}\\\nn
&=&\frac{q_1}{(1-q_1)^2}-\widehat{Q}\frac{1+q_1^2}{(1-q_1)^2}+\widehat{Q}^2\,\frac{1-q_1+q_1^2}{(1-q_1)^2}\\\nn
&=&\widehat{Q}^{2}\Big[\frac{1-q_1+q_1^2}{(1-q_1)^2}-\widehat{Q}^{-1}\frac{1+q_1^2}{(1-q_1)^2}+\widehat{Q}^{-2}\frac{q_1}{(1-q_1)^2}\Big]\\\nn
&=&\widehat{Q}^{2}\,\widehat{Z}^{\,\,\bf
I}_{\tableau{1}\,\tableau{1}}(q_1,\widehat{Q}^{-1})\,. \eea Thus we
see that the two partition functions are equal (up to an overall
factor) if we define the K\"ahler parameters for these two cases,
related by the flop transition, as \bea \widehat{Q}=Q^{-1}\,. \eea
This implies that \bea
W_{\lambda\,\mu}(q_1,N)=q_1^{-\frac{\kappa(\mu)}{2}}\,\Big(Q^{-1}\Big)^{\frac{|\lambda|+|\mu|}{2}}\,\widehat{Z}^{\,\,\bf
II}_{\lambda^{t}\,\mu}(q_1,Q)\,,\,\,\,\,Q=q_1^{-N}\,. \eea Thus we
see that when using the usual topological vertex we get the same
result for the two geometries (with branes) related by flop
transition. This ``symmetry", however, is not preserved by the
refined topological vertex as we will see in the next section.

\section{\sc Refined vertex and Link Homologies}
\label{sec5}
 \vskip 0cm In this section we will determine the
refined open topological string partition functions for the two
configuration of branes on the resolved conifold shown in
\figref{flop}. Let us begin by defining the refined topological
vertex that we will use, \bea\nn
C_{\lambda\,\mu\,\nu}(q_{1},q_{2})&=&\,\Big(\frac{q_{1}}{q_{2}}\Big)^{\frac{||\mu||^2-|\mu|}{2}}\,q_{2}^{\frac{\kappa(\mu)}{2}}\,
q_{2}^{\frac{||\nu||^2}{2}}\,\widetilde{Z}_{\nu}(q_{1},q_{2})\\\nn
&&\times
\sum_{\eta}\,\Big(\frac{q_{2}}{q_{1}}\Big)^{\frac{|\eta|+|\lambda|-|\mu|}{2}}\,
s_{\lambda^{t}/\eta}(q_{1}^{-\rho}\,q_{2}^{-\nu})\,s_{\mu/\eta}(q_{2}^{-\rho}\,q_{1}^{-\nu^{t}})\,
\eea The above definition of the refined topological vertex differs
from the refined vertex in \cite{IKV} by a factor which does not
affect the closed string calculations because it cancels due to
interchanging of $q_{1},q_{2}$ in gluing the vertex along an
internal line. For the open string partition functions this factor
only appears as an over all factor multiplying the partition
function.

The open string refined partition function of the geometry shown in
\figref{flop}(b) is given by \bea
Z_{\lambda\,\mu}(q_{1},q_{2},Q)&=&\sum_{\nu}(-Q)^{|\nu|}\,C_{\varnothing\,\mu\,\nu}(q_{1},q_{2})\,C_{\lambda\,\varnothing\,\nu^{t}}(q_{2},q_{1})\,.
\eea Since \bea
C_{\varnothing\,\mu\,\nu}(q_{1},q_{2})&=&\,\Big(\frac{q_{1}}{q_{2}}\Big)^{\frac{||\mu||^2}{2}}\,q_{2}^{\frac{\kappa(\mu)}{2}}\,
q_{2}^{\frac{||\nu||^2}{2}}\,\widetilde{Z}_{\nu}(q_{1},q_{2})\,s_{\mu}(q_{2}^{-\rho}\,q_{1}^{-\nu^{t}})\\\nn
C_{\lambda\,\varnothing\,\nu^{t}}(q_{2},q_{1})&=&\,\Big(\frac{q_{2}}{q_{1}}\Big)^{\frac{|\lambda|}{2}}\,q_{1}^{\frac{||\nu^{t}||^{2}}{2}}\,
\widetilde{Z}_{\nu^{t}}(q_{2},q_{1})\,s_{\lambda^{t}}(q_{2}^{-\rho}\,q_{1}^{-\nu^{t}})\,,
\eea the open string partition function becomes \bea\nn
Z_{\lambda\,\mu}(q_{1},q_{2},Q)&=&h_{\lambda\,\mu}(q_{1},q_{2})\,\sum_{\nu}(-Q)^{|\nu|}\,q_{2}^{\frac{||\nu||^2}{2}}\,
q_{1}^{\frac{||\nu^{t}||^{2}}{2}}\,\widetilde{Z}_{\nu}(q_{1},q_{2})\,\widetilde{Z}_{\nu^{t}}(q_{2},q_{1})\,
s_{\lambda^{t}}(q_{2}^{-\rho}\,q_{1}^{-\nu^{t}})\,s_{\mu}(q_{2}^{-\rho}\,q_{1}^{-\nu^{t}})\,\\\nn
h_{\lambda\,\mu}(q_{1},q_{2})&=&\,\Big(\frac{q_{1}}{q_{2}}\Big)^{\frac{||\mu||^2}{2}-\frac{|\lambda|}{2}}\,q_{2}^{\frac{\kappa(\mu)}{2}}\,.
\eea The normalized partition function is given by \bea
\widehat{Z}_{\lambda\,\mu}(q_{1},q_{2},Q)=\frac{Z_{\lambda\,\mu}(q_{1},q_{2},Q)}{Z_{\varnothing\,\varnothing}(q_{1},q_{2},Q)}\,
\eea where \bea
Z_{\varnothing\,\varnothing}(q_{1},q_{2},Q)&=&\sum_{\nu}(-Q)^{|\nu|}q_{2}^{\frac{||\nu||^2}{2}}\,q_{1}^{\frac{||\nu^{t}||^{2}}{2}}\,
\widetilde{Z}_{\nu}(q_{1},q_{2})\,\widetilde{Z}_{\nu^{t}}(q_{2},q_{1})\\\nn
&=&\prod_{i,j=1}^{\infty}(1-Q\,q_{1}^{i-\frac{1}{2}}\,q_{2}^{j-\frac{1}{2}})\,.
\eea Recall that the $sl(N)$ Hopf link invariant is related to the
open string partition function as \bea
W_{\lambda\,\mu}(q,N)=q^{-\frac{\kappa(\mu)}{2}}\,\Big(Q^{-1}\Big)^{\frac{|\lambda|+|\mu|}{2}}\,\widehat{Z}^{\,\,\bf
II}_{\lambda^t\,\mu}(q,Q=q^{-N}) \eea The factor
$q^{-\frac{\kappa(\mu)}{2}}$ is the framing factor for the usual
topological vertex. For the case of the refined vertex the framing
factor is given by \cite{IKV}
\bea
f_{\lambda}(q_{1},q_{2})=\Big(\frac{q_{2}}{q_{1}}\Big)^{\frac{||\mu^t||^2-|\mu|}{2}}\,q_{1}^{-\frac{\kappa(\mu)}{2}}
\eea
Therefore we conjecture the following relation between the
homological $sl(N)$ invariants of the Hopf link and the refined open
string partition function\footnote{The factor
$\Big(\frac{q_{1}}{q_{2}}\Big)^{|\lambda|}$ has been introduced to
make the expression symmetric in $\lambda$ and $\mu$.}
\bea
\label{hopflm}
\overline{{\cal P}}_{\lambda\,\mu}(q,t,{\bf a})
&=&\,(-1)^{|\lambda|+|\mu|}\,\Big(\frac{q_{1}}{q_{2}}\Big)^{|\lambda|+|\lambda|\,|\mu|}\,f_{\lambda}(q_{1},q_{2})\,
\Big(Q^{-1}\sqrt{\frac{q_{1}}{q_{2}}}\Big)^{\frac{|\lambda|+|\mu|}{2}}\,\widehat{Z}^{\,\,\bf
II}_{\lambda^t\,\mu}(q_{1},q_{2},Q)\\\nn
&=&\,\Big[\sum_{\nu}(-Q)^{|\nu|}\,q_{2}^{\frac{||\nu||^2}{2}}\,
q_{1}^{\frac{||\nu^{t}||^{2}}{2}}\,\widetilde{Z}_{\nu}(q_{1},q_{2})\,\widetilde{Z}_{\nu^{t}}(q_{2},q_{1})\,
s_{\lambda}(q_{2}^{-\rho}\,q_{1}^{-\nu^{t}})\,s_{\mu}(q_{2}^{-\rho}\,q_{1}^{-\nu^{t}})\Big]\\\nn
&&\times
\Big[Z_{\varnothing\,\varnothing}(q_{1},q_{2},Q)\Big]^{-1}\times
\Big(Q^{-1}\sqrt{\frac{q_1}{q_2}}\Big)^{\frac{|\lambda|+|\mu|}{2}}\times\Big(\frac{q_1}{q_2}\Big)^{|\lambda||\mu|}
(-1)^{|\lambda|+|\mu|}\,.\eea
This is one of the main results of the present paper.
The map between the knot theory
parameters $(q,t,{\bf a})$ and the vertex parameters
$(q_{1},q_{2},Q)$ is given by (\ref{resultt})
where ${\bf a}=q^{N}$, and the limit in which we
recover the usual topological vertex calculation is given by $t=-1$.

\subsection{\sc Unknot}
{}From now on we will drop the superscript ${\bf II}$ on the normalized partition
function and will just write it as $\widehat{Z}_{\lambda\,\mu}(q_{1},q_{2},Q)$.
Below we compute the Poincar\'e polynomial \eqref{hopflm} of the triply-graded
homology for small representations $(\lambda,\mu)$ and compare with known results,
whenever they are available.

For the case $(\lambda,\mu)=(\tableau{1}\,,\varnothing)$ we get
\bea\nn
\overline{{\cal P}}_{\tableau{1}\,\,\varnothing}(t,q,{\bf a})
&=&\,-{\bf a}\frac{\sum_{\nu}(-Q)^{|\nu|}q_{2}^{\frac{||\nu||^2}{2}}\,
q_{1}^{\frac{||\nu^{t}||^{2}}{2}}\,\widetilde{Z}_{\nu}(q_{1},q_{2})\,\widetilde{Z}_{\nu^{t}}(q_{2},q_{1})\,
s_{\tableau{1}}(q_{2}^{-\rho}\,q_{1}^{-\nu^{t}})}{\prod_{i,j=1}^{\infty}(1-Q\,q_{1}^{i-\frac{1}{2}}\,q_{2}^{j-\frac{1}{2}})}\\\nn
&=&-{\bf
a}\Big(\frac{\sqrt{q_{2}}}{1-q_{2}}-Q\sqrt{\frac{q_2}{q_1}}\frac{\sqrt{q_{2}}}{1-q_{2}}\Big)\\\nn
&=&{\bf a}\Big(\frac{1}{q-q^{-1}}-\frac{{\bf
a}^{-2}}{q-q^{-1}}\Big)=\frac{{\bf a}-{\bf
a}^{-1}}{q-q^{-1}}\,,\,\,\,\,\,\,{\bf a}=q^{N}\,. \eea
which is exactly the superpolynomial of the unknot \cite{DGR}.

It is interesting to note that for generic representations
the partition function for the unknot depends on both parameters
$q$ and $t$, whose interpretation we are currently investigating \cite{future}.
However, for totally anti-symmetric representations it is expected
to be only a function of $q$ given by \eqref{grassmh}.
Indeed, for $\tableau{1 1}= \Lambda^2$ and $\tableau{1 1 1}=\Lambda^3$
we find:
\bea
\overline{{\cal P}}_{\Lambda^2}(t,q,{\bf a})&=&{\bf a}^2\,\Big(\frac{\,q^4}{(1-q^2)(1-q^4)}
-\frac{{\bf a}^{-2}\,q^4}{(1-q^2)^2}+\frac{{\bf a}^{-4}\,q^{6}}{(1-q^2)(1-q^4)}\Big) \\ \nn
\overline{{\cal P}}_{\Lambda^3}(t,q,{\bf a})&=&{\bf a}^{3}\Big(-\frac{q^9}{(1-q^2)^3\,(1+2q^2+2q^4+q^6)}
+\frac{{\bf a}^{-2}\,q^9}{(1-q^2)^2\,(1-q^4)}-\frac{{\bf a}^{-4}\,q^{11}}{(1-q^2)^2(1-q^4)} \\ \nn
&&+\frac{{\bf a}^{-6}\,q^{15}\,}{(1-q^2)^3\,(1+2q^2+2q^4+q^6)}\Big)
\eea
in complete agreement with \eqref{grassmh}.
Note, for ${\bf a}=q^{N}$ the partition functions reduce
to a finite polynomials in $q$ with non-negative integer coefficients.

For representations other than the antisymmetric ones
the refined partition function \eqref{hopflm} depends on $t$ in a non-trivial way.

\subsection{\sc Hopf Link}

Let us now consider the Hopf link colored by
$(R_{1},R_{2})=(\tableau{1},\tableau{1})$.
In this case, from Eqs. (\ref{resultt}) and \eqref{hopflm} we get
\bea\nn
\overline{{\cal P}}_{\tableau{1}\,\tableau{1}}(t,q,{\bf
a})&=&\,Q^{-1}\sqrt{\frac{q_{1}}{q_2}}
\frac{\Big(\frac{q_{1}}{q_{2}}\Big)\sum_{\nu}(-Q)^{|\nu|}q_{2}^{\frac{||\nu||^2}{2}}\,q_{1}^{\frac{||\nu^{t}||^{2}}{2}}\,\widetilde{Z}_{\nu}(q_{1},q_{2})\,\widetilde{Z}_{\nu^{t}}(q_{2},q_{1})\,
\Big(s_{\tableau{1}}(q_{2}^{-\rho}\,q_{1}^{-\nu^{t}})\Big)^{2}}{\prod_{i,j=1}^{\infty}(1-Q\,q_{1}^{i-\frac{1}{2}}\,q_{2}^{j-\frac{1}{2}})}\\\nn
&=&{\bf
a}^2\,\Big(\frac{q_{1}}{(1-q_{2})^{2}}-Q\Big(\frac{q_{2}}{q_{1}}\Big)^{\frac{1}{2}}\,
\frac{1+q_{1}-q_{2}+q_{1}q_{2}}{(1-q_{2})^2}+Q^{2}\Big(\frac{q_{2}}{q_{1}}\Big)\,\frac{1-q_{2}+q_{1}q_{2}}{(1-q_{2})^2}\Big)\\\nn
&=&{\bf a}^{2}\Big(\frac{q_{1}}{(1-q_{2})^{2}}-{\bf a}^{-2}\,
\frac{1+q_{1}-q_{2}+q_{1}q_{2}}{(1-q_{2})^2}+{\bf
a}^{-4}\,\frac{1-q_{2}+q_{1}q_{2}}{(1-q_{2})^2}\Big)\\\nn &=&{\bf
a}^{-2}\,\Big(\frac{1-q^2+q^4\,t^2}{(1-q^2)^2}-{\bf
a}^{2}\frac{1+q^2\,t^2-q^2+q^4\,t^2}{(1-q^2)^2}+{\bf
a}^4\frac{q^2\,t^{2}}{(1-q^2)^2}\Big)
\eea
This result agrees with the superpolynomial of the Hopf link computed in Eq.\eqref{khhopfrefined}.

For ${\bf a}=q^{N}$ we get
\bea\nn
\overline{{\cal P}}_{\tableau{1}\,\,\tableau{1}}\Big(q,t,{\bf
a}=q^{N}\Big)&=&
\,q^{-2N}\,\Big\{\frac{1-q^2+q^{4}\,t^2\,}{(1-q^2)^2}-\,q^{2N}\,\frac{1+q^2\,t^2-q^2+q^{4}\,t^2}{(1-q^2)^2}+\,q^{4N}\,\frac{q^2\,t^2}{(1-q^2)^2}\Big\}\\\nn
&=&\,q^{-2N}\,\Big\{\frac{(1-q^2)(1-q^{2N})}{(1-q^2)^2}+t^2\Big(\frac{q^4-q^{2N+2}-q^{2N+4}+q^{4N+2}}{(1-q^2)^2}\Big)\Big\}\\\nn
&=&\,q^{-2N}\,\Big\{q^{N-1}\Big(\frac{q^{N}-q^{-N}}{q-q^{-1}}\Big)+t^2\Big(\frac{q^2-q^{2N}-q^{2N+2}+q^{4N}}{(q-q^{-1})^2}\Big)\Big\}\\\nn
&=&\,q^{-2N}\,\Big\{q^{N-1}\Big(\frac{q^{N}-q^{-N}}{q-q^{-1}}\Big)+t^2\Big(\frac{(1-q^{2N})^2-(1-q^2)(1-q^{2N})}{(q-q^{-1})^2}\Big)\Big\}\\\nn
&=&\,q^{-2N}\,\Big\{q^{N-1}\Big(\frac{q^{N}-q^{-N}}{q-q^{-1}}\Big)+t^2\,\Big(\frac{1-q^{2N}}{q-q^{-1}}\Big)^2+t^2\,q\frac{1-q^{2N}}{q-q^{-1}}\Big\}\\\nn
&=&q^{-2N}\,\Big\{q^{N-1}\Big(\frac{q^{N}-q^{-N}}{q-q^{-1}}\Big)+t^2\,q^{2N}\Big(\frac{q^{N}-q^{-N}}{q-q^{-1}}\Big)^2-t^2\,q^{N+1}\Big(\frac{q^N-q^{-N}}{q-q^{-1}}\Big)\Big\}\\\nn
&=&\,q^{-2N}\,Kh_{N} (2_1^2) \eea
which is {\it exactly} the expression Eq(\ref{khhopfn}) calculated in section 3.

{\bf Hopf link colored by $(\tableau{1},\tableau{1 1})$:}\\ For the
Hopf link colred by $(\tableau{1},\tableau{1 1})$ we get \bea\nn
\overline{{\cal P}}_{(\tableau{1},\tableau{1 1})}(t,q,{\bf a})&=&\frac{{\bf
a}^{-3}(1-q^4+q^{6}\,t^2)}{(1-q^2)^2(1-q^4)}- \frac{{\bf
a}^{-1}\,q^{-2}(1+q^2-q^4-q^6+q^4\,t^2+q^6\,t^2+q^{8}\,t^2)}{(1-q^2)^2(1-q^4)}\\\nn
&&+\frac{{\bf
a}\,q^{-2}(1-q^4+q\,t^2+q^4t^2+q^6\,t^2)}{(1-q^2)^2(1-q^4)}-\frac{{\bf
a}^3\, t^2}{(1-q^2)^2(1-q^4)}\eea There is no knot theory result
with which we can compare this result. However, note that this has
all the right properties. It vanishes for ${\bf a}=1$ i.e., $N=0$
and for ${\bf a}=q^{N}$ it gives $q^{-3N}$ times a finite polynomial
with positive integer coefficients.
\bea
\overline{{\cal P}}_{(\tableau{1},\tableau{1 1})}(t,q,{\bf a}=1)
&=&\overline{{\cal P}}_{(\tableau{1},\tableau{1 1})}(t,q,{\bf a}=q)=0,
\\\nn
\overline{{\cal P}}_{(\tableau{1},\tableau{1 1})}(t,q,{\bf a}=q^2)&=&q^{-6}(1+q^2),
\\\nn \overline{{\cal P}}_{(\tableau{1},\tableau{1 1})}(t,q,{\bf a}=q^3)
&=&q^{-9}(1+2q^2+2q^4+q^6+t^2\,q^6+t^2\,q^8+t^2\,q^{10}),
\\\nn
\overline{{\cal P}}_{(\tableau{1},\tableau{1 1})}(t,q,{\bf a}
=q^{4})&=&q^{-12}(1+2q^2+3q^4+3q^6+2q^8+q^{10}+t^2q^6(1+2q^2+3q^4\\\nn
&&+3q^6+2q^8+q^{10})),\\\nn
\overline{{\cal P}}_{(\tableau{1},\tableau{1 1})}(t,q,{\bf a}=q^5)
&=&q^{-15}(1+2 q^2+3 q^4+\left(4+t^2\right) q^6+\left(4+2 t^2\right)
q^8+\left(3+4 t^2\right) q^{10}\\\nn&&+\left(2+5 t^2\right)
q^{12}+\left(1+6 t^2\right) q^{14}+5 t^2 q^{16}+4 t^2
q^{18}+2 t^2 q^{20}+t^2 q^{22})\\\nn
\overline{{\cal P}}_{(\tableau{1},\tableau{1 1})}(t,q,{\bf a}=q^{6})
&=&q^{-18}(1+2 q^2+3 q^4+\left(4+t^2\right) q^6+\left(5+2 t^2\right) q^8+\left(5+4 t^2\right)
q^{10}+\left(4+6 t^2\right) q^{12}\\\nn &&+\left(3+8 t^2\right)
q^{14}+\left(2+9 t^2\right) q^{16}+\left(1+9 t^2\right) q^{18}+8 t^2
q^{20}+6 t^2 q^{22}+4 t^2 q^{24}\\\nn &&+2 t^2 q^{26}+t^2
q^{28})\eea

\section*{\sc Acknowledgement}

We would like to thank C.~Doran, J.~Rasmussen, and B.~Webster
for valuable discussions.
It is our pleasure to thank the Stony Brook physics department and
the $4^{\rm th}$ Simons Workshop in Mathematics and Physics for hospitality
during the initial stages of this work.
In addition, C.V. thanks the CTP at MIT for hospitality during his sabbatical leave.
The work of S.G. is supported in part by DOE grant DE-FG03-92-ER40701,
in part by RFBR grant 04-02-16880,
and in part by the grant for support of scientific schools NSh-8004.2006.2.
The work of C.V. is supported in part by NSF grants PHY-0244821 and DMS-0244464.

\appendix
\section{\sc Appendix: Other representations}
\label{appendix} In this appendix we write the normalized
partition function of the Hopf link and unknot colored by
other representations of $sl(N)$ which, as usual, we label
by partitions (or Young diagrams). Specifically, we list
simple examples where Young diagrams have at most two columns.

Let us define
\bea\nn
G_{\lambda\,\mu}(-Q,q,t)&:=&\Big[\sum_{\nu}(-Q)^{|\nu|}\,q_{2}^{\frac{||\nu||^2}{2}}\,
q_{1}^{\frac{||\nu^{t}||^{2}}{2}}\,\widetilde{Z}_{\nu}(q_{1},q_{2})\,\widetilde{Z}_{\nu^{t}}(q_{2},q_{1})\,
s_{\lambda}(q_{2}^{-\rho}\,q_{1}^{-\nu^{t}})\,s_{\mu}(q_{2}^{-\rho}\,q_{1}^{-\nu^{t}})\Big]\\\nn
&&\times
\Big[Z_{\varnothing\,\varnothing}(q_{1},q_{2},Q)\Big]^{-1}
\eea
Where we used the identification $(q_{1},q_{2})=(t^2\,q^2,q^2)$ to write $G$ as a function of $q$ and $t$.
In terms of $G_{\lambda\,\mu}(Q,q,t)$ the normalized partition function is given by
\bea
\widehat{Z}_{\lambda\,\mu}(q_{1},q_{2},Q)=\,h_{\lambda\,\mu}\,G_{\lambda^{t}\,\mu}(-Q,q,t)\,,\\\nn
h_{\lambda\,\mu}=q^{\kappa(\mu)}\,t^{||\mu||^2-|\lambda|}\,.
\eea
In the next two sections we list $G_{\lambda\,\mu}$ for various Young diagrams.

\subsection{\sc Unknot}

\fbox{\beqa
\textstyle{G_{(1)}(Q,q,t)=\frac{q}{1-q^2}+\frac{(\frac{q}{t})Q}{1-q^2}}\eeqa}\\
\fbox{\begin{Beqnarray*} &&\textstyle{G_{(1^2)}(Q,q,t)=\frac{q^4}{(1-q^2)(1-q^4)}+\frac{(\frac{q^4}{t})Q}{(1-q^2)^2}+
\frac{(\frac{q^6}{t^2})Q^2}{(1-q^2)(1-q^4)}}\\\nn
&&\textstyle{G_{(2)}(Q,q,t)=\frac{q^2}{(1-q^2)(1-q^4)}+\frac{(1-q^2+q^2\,t^2)Q}{t^3(1-q^2)^2}+
\frac{(1-q^4+q^4\,t^2)Q^2}{t^4\,(1-q^2)(1-q^4)}}\end{Beqnarray*}}\\
\fbox{\beqa\nn \textstyle{G_{(1^3)}(Q,q,t)}&=&\textstyle{\frac{q^9}{(1-q^2)^3\,(1+2q^2+2q^4+q^6)}+\frac{q^9\,Q}{t(1-q^2)^2\,(1-q^4)}+\frac{q^{11}\,Q^2}{t^2(1-q^2)^2(1-q^4)}
+\frac{q^{15}\,Q^3}{t^3(1-q^2)^3\,(1+2q^2+2q^4+q^6)}}\\
\nn \textstyle{G_{(2^{1}\,1^{1})}(Q,q,t)}&=&\textstyle{
\frac{q^5}{(1-q^2)^3\,(1+q^2+q^4)}+\frac{Q(q^3-q^5+q^5\,t^2)}{t^3(1-q^2)^3}+\frac{Q^2(q^3-q^7+q^7\,t^2)}{t^4\,(1-q^2)^3}
+\frac{Q^3(q^5-q^{11}+q^{11}\,t^2)}{t^5\,(1-q^2)^3\,(1+q^2+q^4)}}\eeqa}
\fbox{\beqa\nn \textstyle{G_{(1^4)}(Q,q,t)}&=&\textstyle{\frac{q^{16}}{(1-q^2)^2\,(1-q^4)^2\,(1+q^2+2q^4+q^6+q^8)}+\frac{q^{16}\,Q}{t\,(1-q^2)^4\,(1+2q^2+2q^4+q^6)}
+\frac{q^{18}Q^2}{t^2\,(1-q^2)^2\,(1-q^4)^2}}\\\nn
&&+\textstyle{\frac{q^{22}Q^3}{t^3\,(1-q^2)^4\,(1+2q^2+2q^4+q^6)}+\frac{q^{28}Q^4}{t^4\,(1-q^2)^2(1-q^4)^2\,(1+q^2+2q^4+q^6+q^8)}}
\\\nn \textstyle{G_{(2\,1^{2})}(Q,q,t)}&=&\textstyle{\frac{q^{10}}{(1-q^2)(1-q^4)(1-q^8)}+\frac{Q(q^8-q^{10}+q^{10}\,t^2)}{t^3(1-q^2)^3(1-q^4)}
+\frac{Q^2(q^8+q^{10}-q^{14}-q^{16}+q^{12}\,t^2+q^{14}\,t^2+q^{16}t^2)}{t^4(1-q^2)^2(1-q^4)^2}}\\\nn
&&+\textstyle{\frac{Q^3(q^{10}-q^{16}+q^{16}\,t^2)}{t^5\,(1-q^2)^2\,(1-q^4)^2}+\frac{Q^4(q^{14}-q^{22}+q^{22}\,t^2)}{t^6\,(1-q^2)^2(1-q^4)^2}}
\\\nn
\textstyle{G_{(2^2)}(Q,q,t)}&=&\textstyle{\frac{q^8}{(1-q^2)^2(1-q^4)^2(1+q^2+q^4)}+\frac{Q(q^6-q^{12}+q^8\,t^2+q^{12}\,t^2)}{t^3\,(1-q^2)^4(1+2q^2+2q^4+q^6)}
}\\\nn
&&\textstyle{+\frac{Q^2(q^6-q^8-q^{10}+q^{12}+q^6\,t^2+q^8\,t^2-q^{12}\,t^2-q^{14}\,t^2+q^{10}t^4+q^{14}t^4)}{t^6(1-q^2)^2(1-q^4)^2}}\\\nn
&&
\textstyle{+\frac{Q^3(q^6-q^{10}-q^{12}+q^{16}+q^8\,t^2+q^{10}t^2+q^{12}t^2-q^{14}t^2-q^{16}t^2-q^{18}t^2+q^{14}t^4+q^{18}t^4)}{t^7(1-q^2)^4(1+2q^2+2q^4+q^6)}}
\\\nn
&&\textstyle{+\frac{Q^4(q^8-q^{12}-q^{14}+q^{18}+q^{12}t^2+q^{14}t^2-q^{18}t^2-q^{20}t^2+q^{20}t^4)}{t^8(1-q^2)^2(1-q^4)^2(1+q^2+q^4)}}
\eeqa}\\ \fbox{\beqa
 \textstyle{G_{(1^5)}}&=&\textstyle{-\frac{q^{25}}{(1-q^2)(1-q^4)(1-q^8)(1-q^6)(1-q^{10})}+\frac{q^{25}Q}{t(1-q^2)^2(1-q^4)(1-q^8)(1-q^6)}}\\\nn
&&\textstyle{+\frac{q^{27}Q^2}{t^2\,(1-q^2)^2(1-q^4)^2(1-q^6)}
+\frac{q^{31}Q^3}{t^3\,(1-q^2)^2(1-q^4)^2(1-q^6)}}\\\nn
&&\textstyle{+ \frac{q^{37}Q^4}{t^4(1-q^2)^2(1-q^4)(1-q^8)(1-q^6)}+
\frac{q^{45}Q^5}{t^5(1-q^2)(1-q^4)(1-q^8)(1-q^6)(1-q^{10})}}
\\\nn \textstyle{G_{(2\,1^3)}}&=&\textstyle{-\frac{q^{17}}{\left(-1+q^2\right)^5
\left(1+q^2\right) \left(1+q^2+q^4\right)
\left(1+q^2+q^4+q^6+q^8\right)}-\frac{Q \left(q^{15}-q^{17}+q^{17}
t^2\right)}{\left(-1+q^2\right)^5 \left(1+2 q^2+2 q^4+q^6\right)
t^3}}\\\nn &&\textstyle{+\frac{Q^2 \left(-q^{15}+q^{23}-q^{19}
t^2-q^{23} t^2\right)}{\left(-1+q^2\right)^5 \left(1+2 q^2+2
q^4+q^6\right) t^4}-\frac{Q^3
\left(q^{17}+q^{21}-q^{23}-q^{27}+q^{23} t^2+q^{27}
t^2\right)}{\left(-1+q^2\right)^5 \left(1+2 q^2+2 q^4+q^6\right)
t^5}}\\\nn &&\textstyle{+\frac{Q^4 \left(-q^{21}+q^{29}-q^{29}
t^2\right)}{\left(-1+q^2\right)^5 \left(1+2 q^2+2 q^4+q^6\right)
t^6}-\frac{Q^5 \left(q^{27}-q^{37}+q^{37}
t^2\right)}{\left(-1+q^2\right)^5 \left(1+q^2\right)
\left(1+q^2+q^4\right) \left(1+q^2+q^4+q^6+q^8\right) t^7}}\\\nn
\textstyle{G_{(2^2\,
1)}}&=&\textstyle{\frac{q^{13}}{(1-q^2)^2(1-q^4)(1-q^6)(1-q^8)}
+\frac{Q(q^{11}+q^{13}-q^{19}-q^{21}+q^{13}\,t^2+q^{15}t^2+q^{17}t^2+q^{19}t^2+q^{21}t^2)}{t^3\,(1-q^2)^3(1-q^4)^2(1+q^2+2q^4+q^6+q^8)}}\\\nn
&&\textstyle{+\frac{Q^2(q^{11}-q^{15}-q^{17}+q^{21}+t^2(q^{11}+2q^{13}+q^{15}-q^{19}-2q^{21}-q^{23})
+t^4(q^{15}+q^{17}+q^{19}+q^{21}+q^{23}))}{t^6(1-q^2)^2(1-q^4)^2(1-q^6)}}
\\\nn
&&\textstyle{+\frac{Q^3(q^{11}+q^{13}-2q^{17}-2q^{19}+q^{23}+q^{25}+t^2(q^{13}+2q^{15}+2q^{17}+q^{19}-q^{21}-2q^{23}-2q^{25}-q^{27})+t^4(q^{19}+q^{23}+q^{25}+q^{27}))}{t^7(1-q^2)^2(1-q^4)^2(1-q^6)}+}\\\nn
&& \scriptstyle{\frac{Q^4
q^{13}(1+q^{2}+q^{4}-q^{6}-2q^{8}-2q^{10}-q^{12}+q^{14}+q^{16}+q^{18}
+t^2 q^{4}(1+2q^{2}+2q^{4}+2q^{6}-2q^{10}-2q^{12}-2q^{14}
-q^{33})+t^4
q^{12}(1+q^{2}+q^{4}+q^{6}+q^{8}))}{t^8(1-q^2)^3(1-q^4)^2(1+q^2+2q^4+q^6+q^8)}}\\\nn
&&
+\textstyle{\frac{Q^5(q^{17}-q^{23}-q^{25}+q^{31}+t^2(q^{23}+q^{25}-q^{31}-q^{33})+q^{33}t^4)}{(1-q^2)^3(1-q^4)^2(1+q^2+2q^4+q^6+q^8)}}\eeqa}

\fbox{\beqa \textstyle{G_{(1^6)}}&=&\textstyle{\frac{q^{36}}{\left(-1+q^2\right)^6
\left(1+q^2\right)^3 \left(1+q^2+q^4\right)^2 \left(1+2 q^4+q^6+2
q^8+q^{10}+2 q^{12}+q^{16}\right)}+\frac{q^{36} Q}{\left(1-2
q^2+q^6+q^{10}-2 q^{16}+q^{22}+q^{26}-2 q^{30}+q^{32}\right)
t}}\\\nn && \textstyle{+\frac{q^{38} Q^2}{\left(-1+q^2\right)^6
\left(1+q^2\right)^3 \left(1+q^4\right) \left(1+q^2+q^4\right)
t^2}+\frac{q^{42} Q^3}{\left(-1+q^2\right)^6 \left(1+2 q^2+2
q^4+q^6\right)^2 t^3}}\\\nn &&\textstyle{+\frac{q^{48}
Q^4}{\left(-1+q^2\right)^6 \left(1+q^2\right)^3 \left(1+q^4\right)
\left(1+q^2+q^4\right) t^4}+\frac{q^{56} Q^5}{\left(1-2
q^2+q^6+q^{10}-2 q^{16}+q^{22}+q^{26}-2 q^{30}+q^{32}\right)
t^5}}\\\nn && \textstyle{+\frac{q^{66} Q^6}{\left(-1+q^2\right)^6
\left(1+q^2\right)^3 \left(1+q^2+q^4\right)^2 \left(1+2 q^4+q^6+2
q^8+q^{10}+2 q^{12}+q^{16}\right) t^6}}\\\nn
\textstyle{G_{(2\,1^4)}}&=&
\textstyle{\frac{q^{26}}{\left(-1+q^2\right)^6 \left(1+q^2\right)^3
\left(1+q^4\right) \left(1-q^2+q^4\right)
\left(1+q^2+q^4\right)^2}-\frac{Q \left(-q^{24}+q^{26}-q^{26}
t^2\right)}{\left(-1+q^2\right)^6 \left(1+q^2\right)^2
\left(1+q^4\right) \left(1+q^2+q^4\right) t^3}}\\\nn
&&\textstyle{+\frac{Q^2 \left(q^{24}+q^{26}-q^{34}-q^{36}+q^{28}
t^2+q^{30} t^2+q^{32} t^2+q^{34} t^2+q^{36}
t^2\right)}{\left(-1+q^2\right)^6 \left(1+q^2\right)^3 \left(1+q^2+2
q^4+q^6+q^8\right) t^4}-}\\\nn && \textstyle{\frac{Q^3
\left(-q^{26}-q^{28}-q^{30}+q^{36}+q^{38}+q^{40}-q^{32} t^2-q^{34}
t^2-q^{36} t^2-q^{38} t^2-q^{40} t^2\right)}{\left(-1+q^2\right)^6
\left(1+2 q^2+2 q^4+q^6\right)^2 t^5}}\\\nn &&\textstyle{+\frac{Q^4
\left(q^{30}+q^{32}+q^{34}+q^{36}-q^{40}-q^{42}-q^{44}-q^{46}+q^{38}
t^2+q^{40} t^2+q^{42} t^2+q^{44} t^2+q^{46}
t^2\right)}{\left(-1+q^2\right)^6 \left(1+q^2\right)^3 \left(1+q^2+2
q^4+q^6+q^8\right) t^6}-}\\\nn && \textstyle{\frac{Q^5
\left(-q^{36}+q^{46}-q^{46} t^2\right)}{\left(-1+q^2\right)^6
\left(1+q^2\right)^2 \left(1+q^4\right) \left(1+q^2+q^4\right)
t^7}+\frac{Q^6 \left(q^{44}-q^{56}+q^{56}
t^2\right)}{\left(-1+q^2\right)^6 \left(1+q^2\right)^3
\left(1+q^2+q^4\right)^2 \left(1-q^2+2 q^4-q^6+q^8\right) t^8}}\\\nn
\textstyle{G_{(2^2\,1^2)}}&=&\textstyle{\frac{q^{20}}{\left(-1+q^2\right)^6
\left(1+q^2\right)^3 \left(1+q^4\right)
\left(1+q^2+q^4+q^6+q^8\right)}+\frac{Q \left(-q^{18}+q^{28}-q^{20}
t^2-q^{24} t^2-q^{28} t^2\right)}{\left(-1+q^2\right)^6
\left(1+q^2\right)^2 \left(1+q^4\right)
\left(1+q^2+q^4+q^6+q^8\right) t^3}+}\\\nn && \textstyle{\frac{Q^2
\left(q^{18}-q^{20}-q^{26}+q^{28}+q^{18} t^2+q^{20} t^2-q^{28}
t^2-q^{30} t^2+q^{22} t^4+q^{26} t^4+q^{30}
t^4\right)}{\left(-1+q^2\right)^6 \left(1+q^2\right)^3
\left(1+q^4\right) t^6}}\\\nn &&\textstyle{+\frac{Q^3
\left(-q^{18}+q^{20}+q^{26}-q^{28}-q^{20} t^2+q^{30} t^2-q^{26}
t^4+q^{28} t^4-q^{30} t^4\right)}{\left(-1+q^2\right)^6
\left(1+q^2\right)^2 t^7}+}\\\nn && \textstyle{\frac{Q^4
\left(q^{20}+q^{24}-q^{26}-q^{28}-q^{30}-q^{32}+q^{34}+q^{38}+t^2(q^{24}
+q^{26}+q^{28}+q^{30}-q^{34}-q^{36}-q^{38}-q^{40})+t^4(q^{32}+q^{36}
+q^{40})\right)}{\left(-1+q^2\right)^6 \left(1+q^2\right)^3
\left(1+q^4\right) t^8}-}\\\nn && \textstyle{\frac{Q^5 q^{24}
\left(1+q^{4}-q^{8}-q^{10}-q^{12}-q^{14}+q^{18}+q^{22}+t^2 q^{6}(1
+q^{2}+q^{4}+q^{6}+q^{8}-q^{10}-q^{12}-q^{14}-q^{16}-q^{18}) +t^4
q^{16}(1+q^{4}+q^{8})\right)}{\left(-1+q^2\right)^6
\left(1+q^2\right)^2 \left(1+q^2+2 q^4+2 q^6+2
q^8+q^{10}+q^{12}\right) t^9}+}\\\nn && \textstyle{\frac{Q^6
\left(q^{30}-q^{38}-q^{40}+q^{48}+q^{38} t^2+q^{40} t^2-q^{48}
t^2-q^{50} t^2+q^{50} t^4\right)}{\left(-1+q^2\right)^6
\left(1+q^2\right)^3 \left(1+q^2+2 q^4+2 q^6+2
q^8+q^{10}+q^{12}\right) t^{10}}}\\\nn G_{(2^3)}&=&
\textstyle{\frac{q^{18}}{(1-q^2)^6(1+q^2)^3(1+q^4)(1+q^2+q^4)^2}-\frac{Q(q^{16}-q^{18}+q^{20}-q^{22}+q^{18}t^2-q^{20}t^2+q^{22}t^2)}
{(1-q^2)^6(1+q^2)^2(1+q^2+2q^4+q^6+q^8)t^3}}\\\nn
&&+\textstyle{\frac{Q^2(q^{16}-q^{22}-q^{24}+q^{30}+t^2(q^{16}+q^{18}+q^{20}+q^{22}-q^{26}-q^{28}-q^{30}-q^{32})
+t^4(q^{20}+q^{24}+q^{26}+q^{28}+q^{32}))}{(1-q^2)^6(1+q^2)^3(1+q^2+2q^4+q^6+q^8)t^6}}\\\nn
&&\textstyle{+\frac{Q^3q^{18}(1-q^{2}-q^{4}+q^{8}+q^{10}-q^{12}+t^2q^{-2}(1+q^{2}+q^{4}-q^{6}-2q^{8}-2q^{10}-q^{12}+q^{14}+q^{16}+q^{18}))}
{(1-q^2)^6(1+2q^2+2q^4+q^6)^2t^9}}\\\nn&&
+\textstyle{\frac{Q^3q^{20}t^4(1+q^{2}+2q^{4}+q^{6}+q^{8}-q^{10}-q^{12}-2q^{14}-q^{16}-q^{18}+t^2q^{4}(1+q^{4}+q^{6}+q^{8}+q^{12}))}{(1-q^2)^6(1+2q^2+2q^4+q^6)^2t^9}}\\\nn
&&
+\textstyle{\frac{Q^4(q^{18}-q^{22}-q^{24}-q^{26}+q^{28}+q^{30}+q^{32}-q^{36}
+t^2(q^{18}+q^{20}+2q^{22}-q^{26}-3q^{28}-3q^{30}-q^{32}+2q^{36}+q^{38}+q^{40}))}
{(1-q^2)^6(1+q^2)^3(1+q^2+2q^4+q^6+q^8)t^{10}}}\\\nn
&&\textstyle{\frac{Q^4t^4(q^{22}+q^{24}+2q^{26}+2q^{28}+q^{30}-q^{34}-2q^{36}-2q^{38}-q^{40}-q^{42}
+t^2(q^{30}+q^{34}+q^{36}+q^{38}+q^{42}))}{(1-q^2)^6(1+q^2)^3(1+q^2+2q^4+q^6+q^8)t^{10}}}\\\nn
&&-\textstyle{\frac{Q^5q^{20}(1-q^{2}-q^{6}+q^{10}+q^{14}-q^{16}
+t^2q^2(1+q^{4}-q^{6}-q^{8}-q^{10}-q^{12}+q^{14}+q^{18})+t^4q^8(1+q^{4}-q^{10}-q^{14})
+t^6q^{18}(1-q^{2}+q^{4}))}{(1-q^2)^6(1+q^2)^2(1+q^2+2q^4+q^6+q^8)t^{11}}}\\\nn
&& \textstyle{\frac{Q^6q^{24}
(1-q^{4}-q^{6}-q^{8}+q^{10}+q^{12}+q^{14}-q^{18}+t^2q^4(1
+q^{2}+q^{4}-q^{6}-2q^{8}-2
q^{10}-q^{12}+q^{14}+q^{16}+q^{18}))}{\left(-1+q^2\right)^6
\left(1+q^2\right)^3 \left(1+q^4\right) \left(1+q^2+q^4\right)^2
t^{12}}}\\\nn &&+\textstyle{\frac{Q^6 t^4q^{36}((1+q^{2}+q^{4}-q^{8}
-q^{10}-q^{12})+q^{24} t^6)}{\left(-1+q^2\right)^6
\left(1+q^2\right)^3 \left(1+q^4\right) \left(1+q^2+q^4\right)^2
t^{12}}}
 \eeqa}

 \fbox{\beqa
G_{(2^3\,1)}&=&\textstyle{-\frac{q^{25}}{\left(-1+q^2\right)^7
\left(1+q^2\right)^2 \left(1+q^4\right) \left(1+q^2+q^4\right)^2
\left(1+q^2+q^4+q^6+q^8\right)}}\\\nn &&\textstyle{+\frac{Q
\left(q^{23}+q^{25}+q^{27}-q^{33}-q^{35}-q^{37}+q^{25} t^2+q^{27}
t^2+q^{29} t^2+q^{31} t^2+q^{33} t^2+q^{35} t^2+q^{37}
t^2\right)}{\left(-1+q^2\right)^7 \left(1+2 q^2+2 q^4+q^6\right)^2
\left(1+q^2+2 q^4+2 q^6+2 q^8+q^{10}+q^{12}\right) t^3}-}\\\nn &&
\textstyle{\frac{Q^2q^{23} \left(1-q^{8}-q^{10}+q^{18}+t^2(1+q^{2}
+q^{4}+q^{6}+q^{8}-q^{12}-q^{14}-q^{16}-q^{18}-q^{20})+t^4q^4(1
+q^{4}+q^{6}+q^{8}+q^{10}+q^{12}+q^{16})
\right)}{\left(-1+q^2\right)^7 \left(1+q^2\right)^2 \left(1+2 q^2+4
q^4+5 q^6+6 q^8+5 q^{10}+4 q^{12}+2 q^{14}+q^{16}\right)
t^6}+}\\\nn&&
\textstyle{\frac{Q^3(q^{25}-q^{27}-q^{31}+q^{35}+q^{39}-q^{41}+t^2(q^{23}
+q^{25}+q^{27}-q^{31}-2q^{33}-2q^{35}-q^{37}+q^{41}+q^{43}+q^{45}))}
{(1-q^2)^7(1+q^4)(1+2q^2+2q^4+q^6)^2t^9}}\\\nn &&
+\textstyle{\frac{Q^3t^4(q^{25}+q^{27}+2q^{29}+q^{31}+2q^{33}-2q^{39}-q^{41}-2q^{43}-q^{45}-q^{47}
+t^2(q^{31}+q^{35}+q^{37}+q^{39}+q^{41}+q^{43}
+q^{47}))}{(1-q^2)^7(1+q^4)(1+2q^2+2q^4+q^6)^2t^9}}\\\nn &&
-\textstyle{\frac{Q^4q^{25}(1-q^{6}-2 q^{8}+2
q^{14}+q^{16}-q^{22}+t^2(1+q^{2}+2 q^{4}+q^{6}-2 q^{10}-3 q^{12}-3
q^{14}-2 q^{16}+q^{20}+2
q^{22}+q^{24}+q^{26}))}{\left(-1+q^2\right)^7 \left(1+q^4\right)
\left(1+2 q^2+2 q^4+q^6\right)^2 t^{10}}}\\\nn&&
+\textstyle{\frac{Q^4t^4q^{25}(q^{4}+q^{6}+ 2 q^{8}+2 q^{10}+2
q^{12}+q^{14}-q^{18}-2 q^{20}-2 q^{22}-2
q^{24}-q^{26}-q^{28})+t^6(q^{12}+q^{16}+q^{18}+q^{20}+q^{22}+q^{24}
+q^{28}))}{\left(-1+q^2\right)^7 \left(1+q^4\right) \left(1+2 q^2+2
q^4+q^6\right)^2 t^{10}}}\\\nn && + \textstyle{\frac{Q^5
(q^{27}+q^{31}-q^{33}-q^{35}-2 q^{37}-q^{39}+q^{43}+2
q^{45}+q^{47}+q^{49}-q^{51}-q^{55})}{(-1+q^2)^7 (1+q^2)^2 (1+2 q^2+4
q^4+5 q^6+6 q^8+5 q^{10}+4 q^{12}+2 q^{14}+q^{16}) t^{11}}}\\\nn
&&+\textstyle{\frac{Q^5(q^{29} t^2+q^{31} t^2+2 q^{33} t^2+2 q^{35}
t^2+q^{37} t^2-q^{39} t^2-2 q^{41} t^2-4 q^{43} t^2-4 q^{45} t^2-2
q^{47} t^2-q^{49} t^2+ q^{51} t^2+2 q^{53} t^2+2 q^{55}
t^2)}{(-1+q^2)^7 (1+q^2)^2 (1+2 q^2+4 q^4+5 q^6+6 q^8+5 q^{10}+4
q^{12}+2 q^{14}+q^{16}) t^{11}}}\\\nn&&+\textstyle{\frac{Q^5(q^{57}
t^2+q^{59} t^2+q^{35} t^4+q^{37} t^4+2 q^{39} t^4+2 q^{41} t^4+3
q^{43} t^4+q^{45} t^4+q^{47} t^4-q^{49} t^4-q^{51} t^4-3 q^{53}
t^4-2 q^{55} t^4-2 q^{57} t^4)}{(-1+q^2)^7 (1+q^2)^2 (1+2 q^2+4
q^4+5 q^6+6 q^8+5 q^{10}+4 q^{12}+2 q^{14}+q^{16}) t^{11}}}\\\nn
&&-\textstyle{\frac{Q^5(q^{59} t^4-q^{61} t^4+q^{45} t^6+q^{49}
t^6+q^{51} t^6+q^{53} t^6+q^{55} t^6+q^{57} t^6+q^{61}
t^6)}{\left(-1+q^2\right)^7 \left(1+q^2\right)^2 \left(1+2 q^2+4
q^4+5 q^6+6 q^8+5 q^{10}+4 q^{12}+2 q^{14}+q^{16}\right)
t^{11}}}\\\nn && -\textstyle{\frac{Q^6 (q^{31}+q^{33}+q^{35}-2
q^{39}-3 q^{41}-3 q^{43}-q^{45}+q^{47}+3 q^{49}+3 q^{51}+2
q^{53}-q^{57}-q^{59}-q^{61})}{(-1+q^2)^7 (1+2 q^2+2 q^4+q^6)^2
(1+q^2+2 q^4+2 q^6+2 q^8+q^{10}+q^{12}) t^{12}}}
\\\nn&& -\textstyle{\frac{Q^6q^{35}t^2(1+2 q^{2}+3 q^{4}+3
q^{6}+2 q^{8}-q^{10}-4 q^{12}-6 q^{14}- 6 q^{16}-4 q^{18}-q^{20}+2
q^{22}+3 q^{24}+3 q^{26}+2 q^{28}+q^{30})}{\left(-1+q^2\right)^7
\left(1+2 q^2+2 q^4+q^6\right)^2 \left(1+q^2+2 q^4+2 q^6+2
q^8+q^{10}+q^{12}\right) t^{12}}}\\\nn
&&-\textstyle{\frac{Q^6q^{43}t^4(1+2 q^{2}+3 q^{4}+3 q^{6}+3 q^{8}+2
q^{10}-2 q^{14}-3 q^{16}-3 q^{18}-3 q^{20} -2 q^{22}-
q^{24}+t^2(q^{12}+q^{14}+^{16}+q^{18}+q^{20}+q^{22}+q^{24}))}{\left(-1+q^2\right)^7
\left(1+2 q^2+2 q^4+q^6\right)^2 \left(1+q^2+2 q^4+2 q^6+2
q^8+q^{10}+q^{12}\right) t^{12}}}\\\nn && \textstyle{\frac{Q^7
(q^{37}-q^{43}-q^{45}-q^{47}+q^{51}+q^{53}+q^{55}-q^{61}+q^{43}
t^2+q^{45} t^2+q^{47} t^2-q^{51} t^2-2 q^{53} t^2-2 q^{55}
t^2-q^{57} t^2+q^{61} t^2+q^{63} t^2)}{\left(-1+q^2\right)^7
\left(1+2 q^2+2 q^4+q^6\right)^2 (1+q^2+2 q^4+2 q^6+2
q^8+q^{10}+q^{12}) t^{13}}}\\\nn && +\textstyle{\frac{Q^7(q^{65}
t^2+q^{53} t^4+q^{55} t^4+q^{57} t^4-q^{63} t^4-q^{65} t^4-q^{67}
t^4+q^{67} t^6)}{\left(-1+q^2\right)^7 \left(1+2 q^2+2
q^4+q^6\right)^2 (1+q^2+2 q^4+2 q^6+2 q^8+q^{10}+q^{12}) t^{13}}}
 \eeqa}

\subsection{Hopf Link}

\fbox{\beqa\nn
\textstyle{G_{(1)\,\,(1)}}=\frac{q^2}{(1-q^2)^2}-\frac{Q(1-q^2+q^2t^2+q^4t^2)}{t^3(1-q^2)^2}+\frac{Q^2(1-q^2+q^4t^2)}{t^4(1-q^2)^2}
\eeqa}

\fbox{\beqa \nn
&&\textstyle{G_{(1)\,\,(1^2)}=\frac{q^5}{(1-q^2)^2(1-q^4)}-\frac{Q(q^3-q^7+q^5t^2+q^7t^2+q^9t^2)}{t^3(1-q^2)^2(1-q^4)}
+\frac{Q^2(q^3+q^5-q^7-q^9+q^7t^2+q^9t^2+q^{11}t^2)}{t^4(1-q^2)^2(1-q^4)}-\frac{Q^3(q^5-q^9+q^{11}t^2)}{t^5(1-q^2)^2(1-q^4)}}
\eeqa}

\fbox{ \beqa \nn&& \textstyle{G_{(1)\,\,(1^3)}=
\frac{q^{10}}{(1-q^2)^4(1+2q^2+2q^4+q^6)}-\frac{Q(q^8-q^{14}+q^{10}t^2+q^{12}t^2+q^{14}t^2+q^{16}t^2)}{t^3(1-q^2)^4(1+2q^2+2q^4+q^6)}+
\frac{Q^2(q^8-q^{14}+q^{12}t^2+q^{16}t^2)}{t^4(1-q^2)^3(1-q^4)}}\\\nn
&&\textstyle{-\frac{Q^3(q^{10}+q^{12}+q^{14}-q^{16}-q^{18}-q^{20}+q^{16}t^2+q^{18}t^2+q^{20}t^2+q^{22}t^2)}{t^5(1-q^2)^4(1+2q^2+2q^4+q^6)}
+\frac{Q^4(q^{14}-q^{20}+q^{22}t^2)}{t^6(1-q^2)^4(1+2q^2+2q^4+q^6)}}\\\nn
 && G_{(1,
 1)\,\,(1^2)}=\textstyle{\frac{q^8}{(1-q^2)^4(1+q^2)^2}-\frac{Q(q^6-q^{10}+q^8\,t^2+q^{12}\,t^2)}{t^3(1-q^2)^4(1+q^2)}}
\\\nn &&+\textstyle{\frac{Q^2(q^6-q^8-q^{10}+q^{12}+t^2(q^6+2q^8+q^{10}-q^{12}-2q^{14}-q^{16})+
t^4(q^{10}+q^{12}+2q^{14}+q^{16}+q^{18}))}{t^6(1-q^2)^4(1+q^2)^2}}\\\nn
&&-
\textstyle{\frac{Q^3(q^6-q^8-q^{10}+q^{12}+t^2(q^8+q^{10}-q^{14}-q^{16})+t^4(q^{14}+q^{18}))}{t^7(1-q^2)^4(1+q^2)}}
\\\nn
&&+\textstyle{\frac{Q^4(q^8-q^{10}-q^{12}+q^{14}+t^2(q^{12}+q^{14}-q^{16}-q^{18})+q^{20}\,t^4)}{t^8(1-q^2)^4(1+q^2)^2}}
\eeqa}

\fbox{\beqa\nn
&&\textstyle{G_{(1)\,\,(1^4)}=\frac{q^{17}}{(1-q^2)^5(1+q^2)^2(1+q^4)(1+q^2+q^4)}
-\frac{Q(q^{15}-q^{23}+q^{17}t^2+q^{19}t^2+q^{21}t^2+q^{23}t^2+q^{25}t^2)}
{t^3(1-q^2)^5(1+q^2)(1+q^2+2q^4+q^6+q^8)}}\\\nn
&&\textstyle{+\frac{Q^2(q^{15}+q^{17}-q^{23}-q^{25}+q^{19}t^2+q^{21}t^2+q^{23}t^2+q^{25}t^2+q^{27}t^2)}{t^4(1-q^2)^5(1+q^2)^2(1+q^2+q^4)}-\frac{Q^3(q^{17}+q^{19}+q^{21}-q^{25}-q^{27}-q^{29}+q^{23}t^2+q^{25}t^2+q^{27}t^2+q^{29}t^2+q^{31}t^2)}{t^5(1-q^2)^5(1+q^2)^2(1+q^2+q^4)}}\\\nn
&&\textstyle{+\frac{Q^4(q^{21}+q^{23}+q^{25}+q^{27}-q^{29}-q^{31}-q^{33}-q^{35}+t^2(q^{29}+q^{31}+q^{33}+q^{35}+q^{37}))}{t^6(1-q^2)^5(1+q^2)^2(1+q^2+2q^4+q^6+q^8)}
-\frac{Q^5(q^{27}-q^{35}+q^{37}t^2)}{t^7(1-q^2)^5(1+q^2)^2(1+q^4)(1+q^2+q^4)}}\\\nn
&& G_{(1^2)\,\,(1^3)}=\textstyle{\frac{q^{13}}{(1-q^2)^5(1+q^2)^2(1+q^{2}+q^{4})}-\frac{Q(q^{11}+q^{13}-q^{17}-q^{19}
+t^2(q^{13}+q^{15}+q^{17}+q^{19}+q^{21}))}
{t^3(1-q^2)^5(1+q^2)^2(1+q^2+q^4)}}\\\nn &&
+\textstyle{\frac{Q^2(q^{11}-q^{15}-q^{17}+q^{21}+t^2(q^{11}+2q^{13}+2q^{15}+q^{17}-q^{19}-2q^{21}-2q^{23}-q^{25})
+t^4(q^{15}+q^{17}+2q^{19}
+2q^{21}+2q^{23}+q^{25}+q^{27}))}{t^6(1-q^2)^5(1+q^2)^2(1+q^2+q^4)}}\\\nn
&&
-\textstyle{\frac{Q^3(q^{11}+q^{13}-2q^{17}-2q^{19}+q^{23}+q^{25}+t^2(q^{13}+2q^{15}+3q^{17}+2q^{19}-2q^{23}
-3q^{25}-2q^{27}-q^{29})+t^4(q^{19}+q^{21}+2q^{23}+2q^{25}+2q^{27}+2q^{29}+q^{31}))}{
t^7(1-q^2)^5(1+q^2)^2(1+q^2+q^4)}}\\\nn &&+
\textstyle{\frac{Q^4(q^{13}+q^{15}-2q^{19}-2q^{21}+q^{25}+q^{27}+
t^2(q^{17}+2q^{19}+2q^{21}+q^{23}-q^{25}-2q^{27}-2q^{29}-q^{31})+t^4(q^{25}+q^{27}+q^{29}+q^{31}
+q^{33}))}{t^{8}(1-q^2)^5 (1+q^2)^2(1+q^2+q^4)}}\\\nn &&
-\textstyle{\frac{Q^5(q^{17}-q^{21}-q^{23}+q^{27}+t^2(q^{23}+q^{25}-q^{29}-q^{31})+
t^4\,q^{33})}{t^{9}(1-q^2)^5(1+q^2)^2(1+q^2+q^{4})} }\eeqa}

\fbox{\beqa\nn && \textstyle{G_{(1)\,\,(1^5)}=\frac{q^{26}}{1-2q^{2}+q^{6}+q^{10}-2\,q^{16}+q^{22}+q^{26}-2q^{30}+q^{32}}-\frac{Q(q^{24}-q^{34}
+q^{26}t^2+q^{28}t^2+q^{30}t^2+q^{32}t^2+q^{34}t^2+q^{36}t^2)}
{t^3(1-2q^2+q^6+q^{10}-2q^{16}+q^{22}+q^{26}-2q^{30}+q^{32})}}
\\\nn
&&\textstyle{+\frac{Q^2(q^{24}-q^{34}+q^{28}t^2+q^{32}t^2+q^{36}t^2)
}{t^4(1-q^2)^6(1+q^2)^2(1+q^2+2q^4+q^6+q^8)}
-\frac{Q^3(q^{26}-q^{36}+q^{32}t^2+q^{38}t^2)}{t^5(1-q^2)^6
(1+q^2)^2(1+q^2+q^4)}+\frac{Q^4(q^{30}+q^{34}-q^{40}-q^{44}+q^{38}t^2+q^{42}t^2+q^{46}t^2)}{t^6(1-q^2)^6
(1+q^2)^2(1+q^2+2q^4+q^6+q^8)}}\\\nn&&
\textstyle{-\frac{Q^5(q^{36}+q^{38}+q^{40}+q^{42}+q^{44}-q^{46}-q^{48}-q^{50}-q^{52}-q^{54}+q^{46}t^2+q^{48}
t^2+q^{50} t^2+q^{52} t^2+q^{54} t^2 +q^{56}
t^2)}{t^7(1-2q^2+q^6+q^{10}-2q^{16}+q^{22}+q^{26}-2q^{30}+q^{32})}}\\\nn
&&\textstyle{\frac{Q^6(q^{44}-q^{54}+q^{56}t^2)}
{(1-2q^2+q^6+q^{10}-2q^{16}+q^{22}+q^{26}-2q^{30}+q^{32})t^{8}}}
 \\\nn && G_{(1^2)\,\,(1^4)}=\textstyle{\frac{q^{20}}{(1-q^2)^6(1+q^2)^3(1+q^4)(1+q^{2}+q^{4})}-\frac{Q(q^{18}-q^{26}
+q^{20}t^2+q^{24}t^2+q^{28}t^2)}
{t^3(1-q^2)^6(1+q^2)^2(1+q^4)(1+q^2+q^4)}}
\\\nn&&+\textstyle{\frac{Q^2q^{18}(1-q^{6}-q^{8}+q^{14}+t^2(1+2q^{2}+2q^{4}+2q^{6}+q^{8}-q^{10}-2q^{12}
-2q^{14}-2q^{16}-q^{18})+t^4q^{4}(1+q^{2}+2q^{4}+2q^{6}+3q^{8}+2q^{10}+2q^{12}+q^{14}+q^{16}))}
{t^6(1-q^2)^6(1+q^2)^3(1+q^2+2q^4+q^6+q^8)}}\\\nn
&&-\textstyle{\frac{Q^3(q^{18}-q^{24}-q^{26}+q^{32}+t^2(q^{20}+q^{22}
+q^{24}+q^{26}-q^{30}
-q^{32}-q^{34}-q^{36})+t^4(q^{26}+q^{30}+q^{32}+q^{34}+q^{38}))}{t^7(1-q^2)^6
(1+q^2)^2(1+q^2+q^4)}}\\\nn
&&+\textstyle{\frac{Q^4(q^{20}+q^{22}+2q^{24}-q^{28}-3q^{30}-3q^{32}-q^{34}+2q^{38}+q^{40}+q^{42})}{t^8(1-q^2)^6
(1+q^2)^3(1+q^2+2q^4+q^6+q^8)}}\\\nn&& +
\textstyle{\frac{Q^4t^2(q^{24}+2q^{26}+3q^{28}+4q^{30}+3q^{32}+q^{34}-q^{36}-3q^{38}-4q^{40}-3q^{42}-2q^{44}-q^{46}+t^2(q^{32}
+q^{34}+2q^{36}+2q^{38}+3q^{40}+2q^{42}+2q^{44}+q^{46}+q^{48}))}{t^8(1-q^2)^6
(1+q^2)^3(1+q^2+2q^4+q^6+q^8)}}\\\nn &&
-\textstyle{\frac{Q^5(q^{24}+q^{28}-q^{30}-q^{32}-q^{34}-q^{36}+q^{38}+q^{42}+t^2(q^{30}+q^{32}+q^{34}+q^{36}
-q^{40}-q^{42}-q^{44}-q^{46})+t^4(q^{40}+q^{44}+q^{48}))}{t^9(1-q^2)^6(1+q^2)^2(1+q^2+2q^{4}+q^{6}+q^{8})}}\\\nn
&&
+\textstyle{\frac{Q^6(q^{30}-q^{36}-q^{38}+q^{44}+t^2(q^{38}+q^{40}-q^{46}-q^{48})+t^4q^{50})}{t^{10}(1-q^2)^6(1+q^2)^3(1+q^2+2q^4+q^6+q^8)}}
\\\nn && G_{(1^3)\,\,(1^3)}=\textstyle{\frac{q^{18}}{(1-q^2)^6(1+2q^{2}+2q^{4}+q^6)^2}-\frac{Q(q^{16}-q^{22}
+q^{18}t^2+q^{24}t^2)} {t^3(1-q^2)^6(1+q^2)^2(1+q^2+q^4)}}
\\\nn &&+\textstyle{\frac{Q^2(q^{16}-q^{20}-q^{22}+q^{26}+t^2(q^{16}+q^{18}+q^{20}-q^{26}-q^{28}-q^{30})+t^4(q^{20}+q^{24}+q^{26}
+q^{28}+q^{32}))}{t^6(1-q^2)^6(1+q^2)^2(1+q^2+q^4)}}
\\\nn &&-\textstyle{\frac{Q^3(q^{18}-q^{20}-q^{22}+q^{26}+q^{28}-q^{30}+t^2(q^{16}+2q^{18}+2q^{20}-3q^{24}-4q^{26}
-3q^{28}+2q^{32}+2q^{34}+q^{36}))}{t^9(1-q^2)^6(1+2q^2+2q^4+q^6)^2}}\\\nn
&&
+\textstyle{\frac{Q^3t^4q^{18}(1+2q^{2}+4q^{4}+4q^{6}+4q^{8}+q^{10}-q^{12}-4q^{14}-4q^{16}-4q^{18}
-2q^{20}-q^{22}+t^2q^6(1+q^{2}+2q^{4}+3q^{6}+3q^{8}+3q^{10}+3q^{12}+2q^{14}+q^{16}+q^{18}))}
{t^9(1-q^2)^6(1+2q^2+2q^4+q^6)^2}}\\\nn &&+
\textstyle{\frac{Q^4(q^{18}-q^{20}-q^{22}+q^{26}+q^{28}-q^{30}+
t^2(q^{18}+q^{20}+q^{22}-q^{24}-2q^{26}-2q^{28}-q^{30}+q^{32}+q^{34}+q^{36}))}{t^{10}(1-q^2)^6
(1+q^2)^2(1+q^2+q^4)}}\\\nn
&&+\textstyle{\frac{Q^4t^4(q^{22}+q^{24}+2q^{26}+q^{28}
+q^{30}-q^{32}-q^{34}-3q^{36}
-q^{38}-q^{40}+t^2(q^{30}+q^{34}+q^{36}+q^{38}+q^{42}))}{t^{10}(1-q^2)^6
(1+q^2)^2(1+q^2+q^4)}}
\\\nn
&&-\textstyle{\frac{Q^5(q^{20}-q^{22}-q^{24}+q^{28}+q^{30}-q^{32}+t^2(q^{22}+q^{24}-q^{28}-2q^{30}-q^{32}+q^{36}+q^{38})+
t^4(q^{28}+q^{30}+q^{32}-q^{38}-q^{40}-q^{42})+t^6(q^{38}+q^{44}))}{t^{11}(1-q^2)^6(1+q^2)^2(1+q^2+q^{4})}}\\\nn&&
+\textstyle{\frac{Q^6(q^{24}-q^{26}-q^{28}+q^{32}+q^{34}-q^{36}+t^2(q^{28}+q^{30}-2q^{34}-2q^{36}+q^{40}+q^{42})
+t^4(q^{36}+q^{38}+q^{40}-q^{42}-q^{44}-q^{46}+q^{48}))}{t^{12}(1-q^2)^6(1+2q^2+2q^4+q^6)^2}}
\eeqa}
\newpage
\subsection{Specialization to $Q=-t\,q^{-2N}$: Some examples}

In this section we consider the specialization $Q=-t\,q^{-2N}$ for the case of Hopf link colored by $(R_{1},R_{2})=(1,1^2)$,$(1^2,1^2)$ and $(1^3,1^4)$. We see that $G_{\lambda\,\mu}$ after this specialization is (up to an over all factor) a polynomial in $q$ and $t$.\\
\bea\nn G_{(1)\,(1^2)}(Q=-t,q,t)&=&G_{(1)\,(1^2)}(Q=-t\,q^{-2},q,t)=0,
\\\nn G_{(1)\,(1^2)}(Q=-t\,q^{-4},q,t)&=&-q^{-7}\,t^{-2}(1+q^2), \\\nn
G_{(1)\,(1^2)}(Q=-t\,q^{-6},q,t)&=&-q^{-13}\,t^{-2}(1+2q^2+2q^4+q^6+ t^2\,q^6+t^2\,q^8+t^2\,q^{10}),
\\\nn
G_{(1)\,(1^2)}(Q=-t\,q^{-8},q,t)&=&-q^{-19}\,t^{-2}\,(1+2q^2+3q^4+3q^6+2q^8+q^{10}\\\nn
&&+t^2q^6(1+2q^2+3q^4+3q^6+2q^8+q^{10})),\\\nn
G_{(1)\,(1^2)}(Q=-t\,q^{-10},q,t)&=&-q^{-25}\,t^{-2}(1+2 q^2+3 q^4+\left(4+t^2\right) q^6+\left(4+2
t^2\right) q^8+\left(3+4 t^2\right) q^{10}\\\nn
&&+\left(2+5 t^2\right)
q^{12}+\left(1+6 t^2\right) q^{14}+5 t^2 q^{16}+4 t^2
q^{18}+2 t^2 q^{20}+t^2 q^{22})\\\nn
  G_{(1)\,(1^2)}(Q=-t\,q^{-12},q,t)&=&-q^{-32}\,t^{-2}(1+2 q^2+3 q^4+\left(4+t^2\right) q^6+\left(5+2 t^2\right) q^8+\left(5+4 t^2\right)
q^{10}\\\nn
&&+\left(4+6 t^2\right) q^{12}+\left(3+8 t^2\right)
q^{14}+\left(2+9 t^2\right) q^{16}\\\nn&&
+\left(1+9 t^2\right) q^{18}+8 t^2
q^{20}+6 t^2 q^{22}+4 t^2 q^{24}\\\nn &&+2 t^2 q^{26}+t^2 q^{28}).
\eea

\bea\nn G_{(1^2)\,(1^2)}(Q=-t\,q^{-2N}\,q,t)&=&0\,,\,\,N=0,1\\\nn
G_{(1^2)\,(1^2)}(Q=-t\,q^{-4},q,t)&=&q^{-8}\,t^{-4}\\\nn
G_{(1^2)\,(1^2)}((Q=-t\,q^{-6},q,t)&=&q^{-16}\,t^{-4}(1+q^2+\left(1+t^2\right) q^4+2 t^2 q^6+2 t^2
q^8+t^2 q^{10})\\\nn
 G_{(1^2)\,(1^2)}(Q=-t\,q^{-8},q,t)&=&q^{-24}t^{-4}(1+q^2+\left(2+t^2\right) q^4+\left(1+3 t^2\right) q^6+\left(1+5 t^2\right)
q^8+ 6 t^2 q^{10}\\\nn
&&+\left(5 t^2+t^4\right) q^{12}+\left(3
t^2+t^4\right) q^{14}+ \left(t^2+2 t^4\right) q^{16}+t^4 q^{18}+t^4
q^{20})\\\nn G_{(1^2)\,(1^2)}(Q=-t\,q^{-10},q,t)&=&q^{-32}\,t^{-4}(1+q^2+\left(2+t^2\right)
q^4+\left(2+3 t^2\right) q^6+\left(2+6 t^2\right) q^8\\\nn&&+ \left(1+9
t^2\right) q^{10}+\left(1+11 t^2+t^4\right) q^{12}+\left(11
t^2+2 t^4\right) q^{14}\\\nn&&+ \left(9 t^2+4 t^4\right) q^{16}+\left(6
t^2+5 t^4\right) q^{18}+\left(3 t^2+6 t^4\right)
q^{20}\\\nn&&+\left(t^2+5 t^4\right) q^{22}+4 t^4 q^{24}+2 t^4 q^{26}+t^4
q^{28})\\\nn G_{(1^2)\,(1^2)}(Q=-t\,q^{-12},q,t)&=&q^{-40}\,t^{-4}(1+q^2+\left(2+t^2\right)
q^4+\left(2+3 t^2\right) q^6+\left(3+6 t^2\right) q^8\\\nn&&+ \left(2+10
t^2\right) q^{10}+\left(2+14 t^2+t^4\right) q^{12}+\left(1+17
t^2+2 t^4\right) q^{14}\\\nn&&+ \left(1+18 t^2+5 t^4\right)
q^{16}\\\nn&&+t^2 \left(17+7 t^2\right) q^{18}+t^2 \left(14+11
t^2\right) q^{20}+2 t^2 \left(5+6 t^2\right) q^{22}\\\nn
&&+2 t^2 \left(3+7
t^2\right) q^{24}+3 t^2 \left(1+4 t^2\right) q^{26}+
\left(t^2+11 t^4\right) q^{28}+7 t^4 q^{30}\\\nn
&&+5 t^4 q^{32}+2 t^4
q^{34}+t^4 q^{36})\\\nn
 G_{(1^2)\,(1^2)}(Q=-t\,q^{-14},q,t)&=&q^{-48}\,t^{-4}(1+q^2+\left(2+t^2\right) q^4+\left(2+3 t^2\right) q^6+\left(3+6 t^2\right)
q^8\\\nn&&+\left(3+10 t^2\right) q^{10}\\\nn&&+\left(3+15 t^2+t^4\right)
q^{12}+2 \left(1+10 t^2+t^4\right) q^{14}+ \left(2+24 t^2+5
t^4\right) q^{16}\\\nn&&+\left(1+26 t^2+8 t^4\right) q^{18}+
\left(1+13 t^2 \left(2+t^2\right)\right) q^{20}\\\nn
&&+t^2 \left(24+17
t^2\right) q^{22}+ \left(20 t^2+22 t^4\right) q^{24}+3 t^2
\left(5+8 t^2\right) q^{26}\\\nn&&+2 t^2 \left(5+13 t^2\right) q^{28}+6 t^2
\left(1+4 t^2\right) q^{30}+t^2 \left(3+22 t^2\right)
q^{32}\\\nn&&
+\left(t^2+17 t^4\right) q^{34}+ 13 t^4 q^{36}+8 t^4 q^{38}+5
t^4 q^{40}+2 t^4 q^{42}+t^4 q^{44}) \eea

\bea\nn
G_{(1^3)\,(1^4)}(Q=-t\,q^{-2N},q,t)&=&0\,,\,\,\,N=0,1,2,3\\\nn
G_{(1^3)\,(1^4)}(Q=-t\,q^{-8},q,t)&=&-q^{-19}\,t^{-6}(1+q^2+q^4+q^6)\\\nn
G_{(1^3)\,(1^4)}(Q=-t\,q^{-10},q,t)&=&-q^{-33}\,t^{-6}(1+2 q^2+3
q^4+\left(4+t^2\right) q^6+\left(4+2 t^2\right) q^8+\left(3+4
t^2\right) q^{10}\\\nn&&+\left(2+5 t^2\right) q^{12}+\left(1+6
t^2\right) q^{14}+ 5 t^2 q^{16}+4 t^2 q^{18}+2 t^2 q^{20}+t^2
q^{22})\\\nn
G_{(1^3)\,(1^4)}(Q=-t\,q^{-12},q,t)&=&-q^{-47}\,t^{-6}(1+2 q^2+4
q^4+\left(6+t^2\right) q^6+\left(8+3 t^2\right) q^8 +\left(9+7
t^2\right) q^{10}\\\nn&&+ \left(9+12 t^2\right) q^{12}+\left(8+18
t^2\right) q^{14}+\left(6+23 t^2+t^4\right) q^{16}\\\nn&& +
\left(4+26 t^2+2 t^4\right) q^{18}+\left(2+26 t^2+4 t^4\right)
q^{20}+ \left(1+23 t^2+6 t^4\right) q^{22}\\\nn&& +\left(18 t^2+8
t^4\right) q^{24}+\left(12 t^2+9 t^4\right) q^{26}+\left(7 t^2+9
t^4\right) q^{28}\\\nn&&+\left(3 t^2+8 t^4\right) q^{30}+\left(t^2+6
t^4\right) q^{32}+4 t^4 q^{34}+2 t^4 q^{36}+t^4 q^{38})\\\nn
G_{(1^3)\,(1^4)}(Q=-t\,q^{-14},q,t)&=&-q^{-61}\,t^{-6}(1+2 q^2+4
q^4+\left(7+t^2\right) q^6 +\left(10+3 t^2\right) q^8+\left(13+8
t^2\right) q^{10}\\\nn&&+ \left(16+15 t^2\right) q^{12}+\left(17+26
t^2\right) q^{14}+\left(17+38 t^2+t^4\right) q^{16}\\\nn&&+
\left(16+52 t^2+3 t^4\right) q^{18}+\left(13+7 t^2
\left(9+t^2\right)\right) q^{20}\\\nn&&+\left(10+72 t^2+13
t^4\right) q^{22}+\left(7+74 t^2+21 t^4\right) q^{24}\\\nn&&+
\left(4+72 t^2+30 t^4\right) q^{26}+\left(2+63 t^2+39 t^4\right)
q^{28}\\\nn&&+\left(1+52 t^2+46 t^4+t^6\right) q^{30} +t^2
\left(38+50 t^2+t^4\right) q^{32}\\\nn&&+2 t^2 \left(13+25
t^2+t^4\right) q^{34}+t^2 \left(15+t^2\right) \left(1+3 t^2\right)
q^{36}\\\nn &&+ t^2 \left(8+39 t^2+4 t^4\right) q^{38}+t^2
\left(3+30 t^2+4 t^4\right) q^{40}\\\nn&&+\left(t^2+21 t^4+5
t^6\right) q^{42}+t^4 \left(13+4 t^2\right) q^{44}+t^4 \left(7+4
t^2\right) q^{46}\\\nn&&+3 t^4 \left(1+t^2\right) q^{48}+\left(t^4+2
t^6\right) q^{50}+t^6 q^{52}+t^6 q^{54}) \eea

\addcontentsline{toc}{section}{\refname}
\bibliography{physics}

\end{document}